\documentclass[preprint,12pt,authoryear]{elsarticle}

\usepackage{amssymb}
\usepackage{amsmath}

\usepackage{amsmath, amsfonts, graphicx}
\usepackage{dsfont, xcolor}
\usepackage{booktabs,caption,subcaption}
\usepackage{natbib}
\usepackage{url}
\usepackage{algorithm,algpseudocode}

\usepackage{adjustbox}

\journal{Computational Statistics \& Data Analysis}

\begin{document}

\begin{frontmatter}



\title{Graph-based Change Point Detection for Functional Data} 


\author[label1]{Jeremy VanderDoes\corref{cor1}}
\author[label1]{Shojaeddin Chenouri} 

\affiliation[label1]{organization={Department of Statistics and Actuarial Science},
            addressline={200 University Avenue West}, 
            city={Waterloo},
            postcode={N2L 3G1}, 
            state={ON},
            country={Canada}}
\cortext[cor1]{Corresponding author, jeremy.vanderdoes@uwaterloo.ca}

\begin{abstract}

Modeling functions that are sequentially observed as functional time series is becoming increasingly common. In such models, it is often crucial to ensure data homogeneity. We investigate the sensitivity of graph-based change point detection for changes in the distribution of functional data that demarcate homogeneous regions. Related test statistics and thresholds for detection are given. A key factor in the efficacy of such tests is the graph construction. Practical considerations for constructing a graph on arbitrary data are explored. Simulation experiments investigate tuning parameters for graph construction and evaluate the graph-based methods in comparison to existing functional methods. 
In addition to sensitivity of lower and higher order changes, robustness to the tuning parameter choices, and practical recommendations, are shown. Applications to multi-year pedestrian counts, high-frequency asset returns, and continuous electricity prices corroborate the simulation results. 
\end{abstract}




\end{frontmatter}




\section{Introduction}

Data collection has become a widespread phenomenon over the last two decades with increased computational power. When investigating the resultant data sets, disruptions to data homogeneity are often of interest to ensure the validity of a model or to identify the effect of some events. Change point analysis encompasses the characterization of changes in ordered data. The seminal work by \cite{page:1954,page:1955} focused on detecting changes in a stream of scalar data as quickly as possible under ongoing data collection; see also \cite{barnard:1959} and \cite{shiryaev:1963}. Conversely, \cite{page:1957} and \cite{picard:1985} conducted change point detection on datasets in which all sample points were collected. Respectively, these two approaches are often coined sequential and retrospective analysis \citep{aminikhanghahi:cook:2017}. 

Change point analysis experienced a renaissance in the 1980s and 1990s; see \cite{worsley:1986} and \cite{james:james:siegmund:1987}. Since then, work related to change point methods on scalar data has become extensive, e.g. see the monographs \cite{basseville:nikiforov:1993}, \cite{brodsky:darkhovsky:1993}, \cite{csorgo:horvath:1997}, and \cite{horvath:rice:2024}. Early change point methods focused on independent Gaussian observations and aimed to detect a single change in the mean, or later covariance, in various settings, e.g. \cite{carlin:gelfand:adrian:1992} and \cite{andrews:1993}. Subsequent methods were developed to consider more general changes, e.g. \cite{huskova:meintanis:2006} consider higher-order changes in independent data. Many single change point detection methods are based on the cumulative sum (CUSUM) statistic presented in \cite{page:1954}. Although originally designed for detecting a single change point, CUSUM statistics have been extended to detect multiple changes. One of the most popular extensions is binary segmentation \citep{scott:knott:1974, vostrikova:1981, venkatraman:1992}. Binary segmentation recursively searches for change points based on sub-segments from previously detected changes. Other multiple change point detection methods also exist, e.g. \cite{yao:1988} and \cite{yao:au:1989} use exhaustive search while \cite{killick:fearnhead:eckley:2012} minimize some cost function. See also the review by \cite{niu:hao:zhang:2016}.

While change point detection in the univariate situation has received significant attention, restricting the dimension of the data to $\mathbb{R}$ can be limiting in many applications; see review by \cite{truong:oudre:vayatis:2020}. That said, change point analysis of multivariate data have received considerable attention in the past 20 years; see \cite{galeano:pena:2007}, \cite{lavielle:teyssiere:2006}, \cite{aue:hormann:horvath:reimherr:2009}, \cite{zhang:etal:2010}, \cite{wang:zou:yin:2018}, \cite{kao:trapani:urga:2018}, \cite{enikeeva:harchaoui:2019}, \cite{chenouri:mozaffari:rice:2020}, and \cite{ramsay:chenouri:2023}. 

Multivariate change point detection methods highlight that CUSUM-based statistics are readily computable but often rely on parametric assumptions or asymptotic properties to detect changes and may perform poorly under certain high-dimensional scenarios; see \cite{galeano:pena:2007}, \cite{lavielle:teyssiere:2006}, \cite{aue:hormann:horvath:reimherr:2009}, \cite{zhang:etal:2010}, \cite{wang:zou:yin:2018}, \cite{kao:trapani:urga:2018}, \cite{enikeeva:harchaoui:2019}, \cite{chenouri:mozaffari:rice:2020}, and \cite{ramsay:chenouri:2023}. To mitigate this, \cite{chen:zhang:2015} presented a graph-based approach. This approach only requires a definition of distances between observations. It has shown efficacy detecting changes in scalar and multivariate settings; see \cite{chen:zhang:2013}, \cite{shi:wu:rao:2017}, \cite{chen:friedman:2017}, \cite{chu:chen:2019}, \cite{chen:2019:dep}, \cite{chen:2019}, \cite{zhang:chen:2021}, \cite{nie:nicolae:2021}, and \cite{zhou:chen:2022}. A recent review by \cite{chu:chen:2022} highlights key developments for graph-based change point estimation.

Fully functional methods to detect higher order changes for functional time series (FTS) are largely missing from the literature. Recent work by \cite{horvath:rice:vanderdoes:2025} consider an approach based on empirical characteristic functionals to detect distributional changes. The graph-based approach proposed by \cite{chen:zhang:2015} is more general than typically presented on multivariate applications and can be directly applied to functional data. However, the lack of functional literature means its performance, and tuning parameter choices, have not been investigated to our knowledge.

In this paper, we fill this gap by investigating the efficacy of the graph-based framework in the context of retrospective functional data change point estimation. An extensive simulation study is conducted to evaluate the size and power of graph-based methods under various scenarios, including the low sample settings. The effect of tuning parameter choices, robustness under parametric and non-parametric assumptions, and performance of graph-based methods are evaluated. Practical recommendations for default values of the tuning parameters are given along with the performance of graph-based methods under ideal and non-ideal selection of the tuning parameters. Performance of the graph-based methods for detecting mean, covariance, and distributional changes is evaluated and compared to existing functional methods. Binary segmentation is used to extend the tests to the multiple change scenario. Several examples of FTS are studied.

The remainder of this paper is organized as follows. The four graph-based change point detection statistics are presented in section \ref{sec:model}, along with their theoretical properties. Approaches for computing thresholds for the statistics are also given. Section \ref{sec:simulations} investigates the proposed test statistics via simulations in retrospective scenarios. The simulations examine performance under various settings such as variety of observed data length, dependence structure, change point types, and location of the change(s). Performance of the graph-based statistics under tuning parameter choices are considered. The graph-based methods are compared to existing functional change point detection methods. Section \ref{sec:applications} applies graph-based change point detection on multi-year pedestrian counts, high-frequency stock returns, and continuous electricity prices. Finally, section \ref{sec:graph_conclusion} provides a discussion on graph-based methods and our contributions. All code and data used in this work can be found at \url{github.com/jrvanderdoes/fungraphs} in the $\mathtt{R}$ package {\tt fungraphs}.

\section{Functional graph-based change point analysis} \label{sec:model}

We consider FTS of the form $\{X_i(t): i=1,\dots,n, \; t\in[0,1]\}$. These observations are often assumed to take value in the Hilbert space $L^2[0,1]$ of real-valued, measurable, and square-integrable functions on the unit interval which is equipped with an inner product. More generally, we can consider data in $L^p[0,1]$ for $p=1,2,\dots$. For $X(t),Y(t)\in L^p[0,1]$, let the corresponding norm be denoted
\begin{equation} \label{eq:lp_norm}
    ||X-Y||_p =\left(\int_0^1 |X(t)-Y(t)|^p dt\right)^{1/p}\,.
\end{equation}
Below we focus on $L^1$ and $L^2$ norms, referred to as simply $L^1$ and $L^2$.

Graph-based change point methods work by treating the data as part of a graph. A graph $G$ can be described as a set of vertices and a set of edges that connect the vertices. Graph-based change point detection treats the observations as graph vertices, connects the observations via edges, and investigates the resultant edge-defined clusters. Edges are defined using a distance metric and a method for selecting edges based on the distances. Intuitively, a change is detected when there is a cluster of strongly connected observations, or vertices with a relatively large number of connecting edges, which are less connected to observations (vertices) outside the cluster. Formally, changes are detected using a two-sample test statistic based on edge-count.

\subsection{Single change analysis} \label{sec:single}

In the framework of the at-most-one-change (AMOC), detecting a change can be framed as a hypothesis testing problem. Let the probability law governing the process $X_i$ be denoted by $F_i$ for $i=1, \dots, n$. We consider the null hypothesis
\begin{equation} \label{eq:H0}
    {H_0}: F_1 = \cdots  = F_n
\end{equation}
against the alternative hypothesis
\begin{align} \label{eq:H1_single}
    \begin{split}
        {H_A}: \mbox{ There exists an integer $k^*=1,\dots, n$ such that } \\
        F_1 = \cdots = F_{k^*} \ne   F_{k^*+1} = \cdots = F_{n}\,.
    \end{split}
\end{align}
Section \ref{sec:multiple} discusses how one may adapt the proposed test to a more general alternative which allows for multiple changes in the sample of FTS.

Graph-based change point detection is based on transforming observations into a graph and analyzing potential clusters. Additional details on the graph construction are contained in section \ref{sec:construction}. Clusters in the graph are determined by an edge-count metric that examines the number of edges connecting observations. Changes in the data are detected by comparing the edge-counts to that which would be expected due to randomness. Previous works in the context of multivariate data have identified four potential statistics, namely: original, weighted, generalized, and max-type edge-count statistics \citep{shi:wu:rao:2017, chen:friedman:2017, chu:chen:2019}. Each statistic splits the observed data into $2$ groups at observation location $k=1, 2, \dots, n$. Let $i$ and $j$ denote observations, $i,j=1,\dots,n$ and $i\neq j$. Denote a potential edge between $i$ and $j$ as $(i,j)$. The test statistics are based on three quantities that measure the edge counts,
\begin{align*}
    R_0(k) &= \sum_{(i,j)\in G} \mathds{I}(g_i(k)\neq g_j(k))\,, \\
    R_1(k) &= \sum_{(i,j)\in G} \mathds{I}(g_i(k)= g_j(k)=0)\,, \quad\text{and} \\
    R_2(k) &= \sum_{(i,j)\in G} \mathds{I}(g_i(k)= g_j(k)=1) \,,
\end{align*}
where $g_i(k)=\mathds{I}(i>k)$, i.e. the indicator function that a given observation $i$ is grouped into the second group. Because $k$ divides the observations into two groups, these quantities can measure the number of edges between the two groups ($R_0(k)$) or within the groups ($R_1(k)$ and $R_2(k)$). If the observations before and after $k$ come from different distributions, the observations in each group should be closer to each other than observations in the other group. Relatively small $R_0(k)$ or large $R_1(k)$ and $R_2(k)$ provide evidence to reject the null in \eqref{eq:H0} and conclude that there is a structural change in the FTS at time $k$. Unfortunately due to the curse of dimensionality, observations in the same group may not appear close together. Measuring the closeness and remoteness of observations can be valuable in mitigating power and size degradation in high dimensional settings. In FTS, this refers to time series that are observed at a high resolution. The proposed graph-based statistics are defined:
\begin{align}
    Z_0(k) &= -\frac{R_0(k)-\text{E}(R_0(k))}{\sqrt{\text{Var}(R_0(k))}}\,,\label{eq:teststat_o}\\
    Z_w(k) &= \frac{R_w(k)-\text{E}(R_w(k))}{\sqrt{\text{Var}(R_w(k))}}\,,\label{eq:teststat_w}\\
    S(k) &= \begin{pmatrix}
                R_1(k)-\text{E}(R_1(k))\\
                R_2(k)-\text{E}(R_2(k))
            \end{pmatrix}^\top 
            \Sigma_R^{-1} 
            \begin{pmatrix}
                R_1(k)-\text{E}(R_1(k))\\
                R_2(k)-\text{E}(R_2(k))
            \end{pmatrix}\,,\label{eq:teststat_g} \quad\text{and}\\
    M(k) &= \max\left(Z_w(k), \left|Z_{\text{diff}}(k)\right|\right)\,,\label{eq:teststat_m}
\end{align}
where
\begin{align*}
    R_w(k) &= \frac{n-k-1}{n-2}R_1(k)+\frac{k-1}{n-2}R_2(k)\,,\\
    R_d(k) &= R_1(k)-R_2(k)\,,\\
    Z_{\text{diff}}(k) &= \frac{R_d(k)-\text{E}(R_d(k))}{\sqrt{\text{Var}(R_d(k))}}\,, \quad \text{ and }\\
    R_d(k) &= R_1(k)-R_2(k)\,.
\end{align*}
We respectively refer to the statistics as the original \eqref{eq:teststat_o}, weighted \eqref{eq:teststat_w}, generalized \eqref{eq:teststat_g}, and max-type \eqref{eq:teststat_m} edge-count statistics.

The original and weighted edge-count statistics are designed to detect mean changes. As discussed for the case of scalar and multivariate data in \cite{chen:chu:2023}, the original statistic tends to perform well when the change point is located near the middle of the data. However, its performance diminishes when the number of observations on either side of the change are uneven, i.e. the change point is near either boundary. This effect is due to inaccurate variance estimation for segments that are smaller, i.e. closer to the boundary. The weighted statistic is computed based on the weighted average of $R_1(k)$ and $R_2(k)$ to accommodate the effect of the sample size in each segment. This leads to improved performance near the end points while maintaining the performance for mid-sample changes.

In the multivariate settings, the generalized and max-type edge-count statistics are recommended for general changes. The generalized and max-type test statistics typically exhibit similar performance to each other \citep{chen:zhang:2015, chen:friedman:2017, chu:chen:2019, chen:chu:2023}. These two statistics are designed to be more robust to high-dimensional data by using both the closeness and remoteness of observations to define the statistics. \cite{chu:chen:2019} show that the generalized edge-count statistic can be decomposed into the uncorrelated quantities $Z_w(k)$ and $Z_{\text{diff}}(k)$, where the first is the weighted statistic that is particularly sensitive to location changes and the second is sensitive to scale changes. The max-type edge-count statistic takes the maximum of the uncorrelated quantities, i.e. the max-type statistic is defined as the maximum of $Z_w(k)$ and $|Z_{\text{diff}}(k)|$.

A change point is proposed at the location where the test statistic takes its maximum value, i.e.,
\begin{align}
    \hat{k}_{\text{0}}^* &=\underset{k \in \{1,\dots,n\} }{\text{argmax}}\text{ } Z_0(k)\,,\label{test-def:1}\\
    \hat{k}_{\text{w}}^* &=\underset{k \in \{1,\dots,n\} }{\text{argmax}}\text{ } Z_w(k)\,,\label{test-def:2}\\
    \hat{k}_{\text{S}}^* &=\underset{k \in \{1,\dots,n\} }{\text{argmax}}\text{ } S(k)\,,\label{test-def:3} \quad\text{and}\\
    \hat{k}_{\text{M}}^* &=\underset{k \in \{1,\dots,n\} }{\text{argmax}}\text{ } M(k)\,.\label{test-def:4}
\end{align}
The smallest value of $k$ is taken in the case of ties. In practice, $k$ is restricted to the interval $n_0,\dots,n_1$ where $1< n_0 \leq n_1 < n$ to prevent change point estimation based on an inadequate sample size. \cite{gseg} suggest using $n_0=0.05n$ and $n_1=0.95n$. We take this approach, but encourage further investigations into this choice.

The proposed change point is deemed significant if the test statistic from \eqref{eq:teststat_o}--\eqref{eq:teststat_m} is at least as big as some properly calibrated threshold. One method to determine the threshold is by using permutation. When the data are assumed to be exchangeable and under the null hypothesis in \eqref{eq:H0}, test statistics computed from permuted data should be similar to the statistic computed from the original data. In practice, random shuffling is used to minimize computation time. 

Let $T_n$ denote a test statistic computed from \eqref{test-def:1}--\eqref{test-def:4} on the original data. For a user specified number $M$, which is often less than the total possible number of permutations, the observed data $(X_1,\dots,X_n)$ is randomly permuted $M$ times, producing permuted samples $(X_1^{(m)},\dots,X_n^{(m)}),$ $m\in \{1,\dots,M\}$. Denoting the test statistics computed from the $m'$th permuted sample $T_n^{(m)}$, the threshold for the test statistic at significance $\alpha\in(0,1)$ is defined as
\begin{equation} \label{threshold}
     t_{n,(1-\alpha)} = \inf\left\{ q \; : \; \frac{1}{M} \sum_{ m=1}^M \mathds{1}( T_n^{(m)} \le q) > 1-\alpha \right\}\,.
\end{equation}
When the underlying data are serially independent, this critical value (for large $M$) will furnish a test with exact size $\alpha$. 

When the data are not exchangeable, the dependency structure can be modeled using a FTS model, e.g. a functional autoregressive moving-average model \citep{aue:norinho:hormann:2015} or the projection approach proposed by \cite{hyndman:ullah:2007}. The residuals of well-fit models are approximately exchangeable, and may be used to generate pseudo-samples. Alternatively, \cite{chen:2019:dep} proposes using circular block permutation for multivariate data to threshold the test statistics; see also block permutation discussions in \cite{carlstein:1986} and \cite{lahiri:1999}. Selection of the bandwidth length in such methods has been a topic of discussion, e.g. see \cite{horvath:rice:whipple:2016} and \cite{rice:shang:2017} for data-driven bandwidth selection.

Permutation tests and random shuffling can become computationally expensive as $n$ becomes large. Hence, analytic formulas to approximate the permutation $p$-values have been proposed. \cite{chen:zhang:2015} and  \cite{chu:chen:2019} propose approximations which are distribution-free and do not depend on the underlying graph $G$ nor the data used to construct the graph. The approximations are based on the fact that the limiting distributions of the test statistics converge to respective Gaussian processes, with known covariance functions. However in this paper, we focus on random shuffling and leave details of approximation $p$-values to other works; see \cite{chen:zhang:2015},  \cite{chu:chen:2019}, and \cite{chen:2019:dep}. 

A proposed change point is deemed significant at a $1-\alpha$ level if $T_n>t_{n,(1-\alpha)}$ where $T_n$ is any of the test statistics defined in \eqref{test-def:1}--\eqref{test-def:4} and $t_{n,(1-\alpha)}$ is the corresponding threshold defined in \eqref{threshold}.

\subsection{Multiple change analysis} \label{sec:multiple}

For many applications, the data appear to exhibit more than one change point. Evaluating data for the presence of multiple change points can again be phrased as a hypothesis testing problem. This problem compares the null hypothesis defined in \eqref{eq:H0} against the $m$ change alternative hypothesis 
\begin{equation} \label{eq:H1_mult}
    H_A: F_1=\dots=F_{k_1^*} \neq F_{k_1^*+1}= \dots = F_{k_2^*} \neq \dots \neq F_{k_m^*+1} = \dots = F_n\,,
\end{equation}
which states there are $m$ changes, $1\leq k_1 < \dots < k_m \leq n$.

We use binary segmentation (BS) to estimate multiple change points. Recall that BS recursively applies a single-change point detection method. If a change is deemed significant, the data are segmented at that point, and the single-change method is reapplied to the resulting sub-segments. For reference, the BS algorithm is outlined in Algorithm~\ref{alg:binseg}. 
\begin{algorithm}
    \caption{{\bf Binary segmentation algorithm.} BINSEG$(\ell,u,\xi_n)$}\label{alg:binseg}
    \begin{algorithmic}
        \State $\ell \gets $ start index; $u\gets $ end index; $\xi_n \gets $ threshold
        
        \State \algorithmicif\ $u-\ell\leq 1$\,, \ \algorithmicthen\ STOP\,, \algorithmicelse\ compute $\hat{k}^*:=\text{argmax}_{\ell<k<u}||T_{\ell,u}(k)||$\,; $T=||T_{\ell,u}^{\hat{k}^*}||$

        \If{$T>t_{n,(1-\alpha)}$}
            \State add $\hat{k}^*$ to the set of estimated change points
            \State \textbf{run} BINSEG$(\ell, \hat{k}^*,t_{n,(1-\alpha)})$ and BINSEG$(\hat{k}^*,u,t_{n,(1-\alpha)})$
            \State \algorithmicelse\ STOP
        \EndIf
    \end{algorithmic}
\end{algorithm}

BINSEG$(\ell,u,\xi_n)$ returns estimates of the number of changes points $\hat{m}$ and their locations $\hat{k}_1^*,\dots,\hat{k}_m^*$.  The consistency of BS for multiple change point detection has been shown under various conditions, e.g. see \cite{scott:knott:1974}, \cite{vostrikova:1981}, \cite{venkatraman:1992}, \cite{zhang:chen:2021}, and \cite{rice:zhang:2022}.

\subsection{Graph construction} \label{sec:construction}

Graph-based change point detection is based on transforming observations into a graph and detecting changes based on clusters of that graph. With the test statistics established, we now discuss how to construct the graphs. Although distributional properties of the test statistics have the same asymptotic properties under very mild restrictions on the graph, the graph construction affects the power of the test statistics. \cite{chen:friedman:2017} discuss the history and assumptions for several popular graphs in the multivariate setting. 

\begin{figure}
    \centering
    \begin{subfigure}{0.30\textwidth}
        \includegraphics[width=\textwidth]{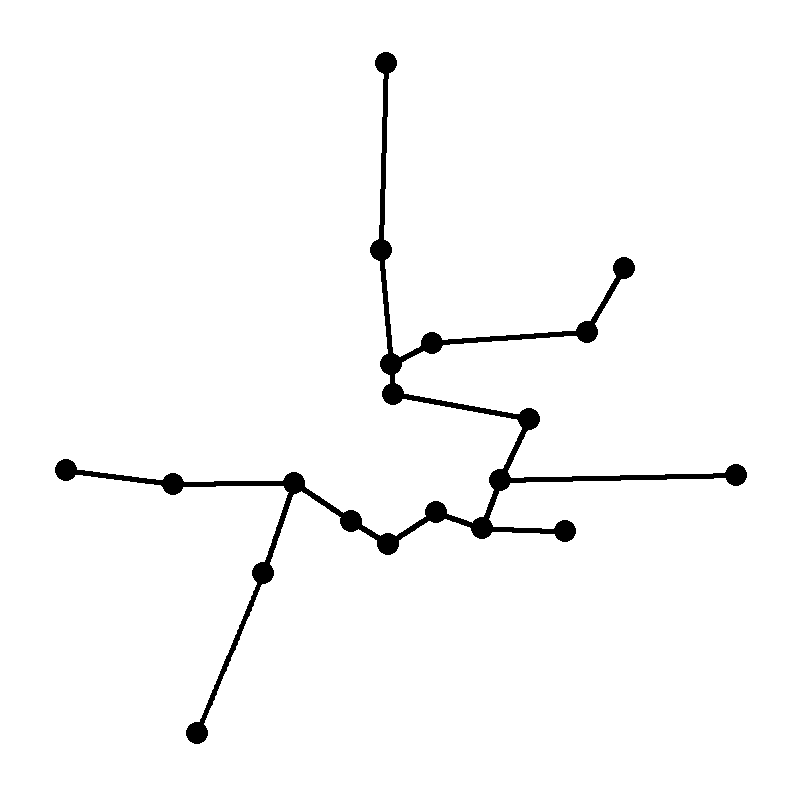}
        \caption{MST graph}
        \label{fig:graphsA}
    \end{subfigure}
    \begin{subfigure}{0.30\textwidth}
        \includegraphics[width=\textwidth]{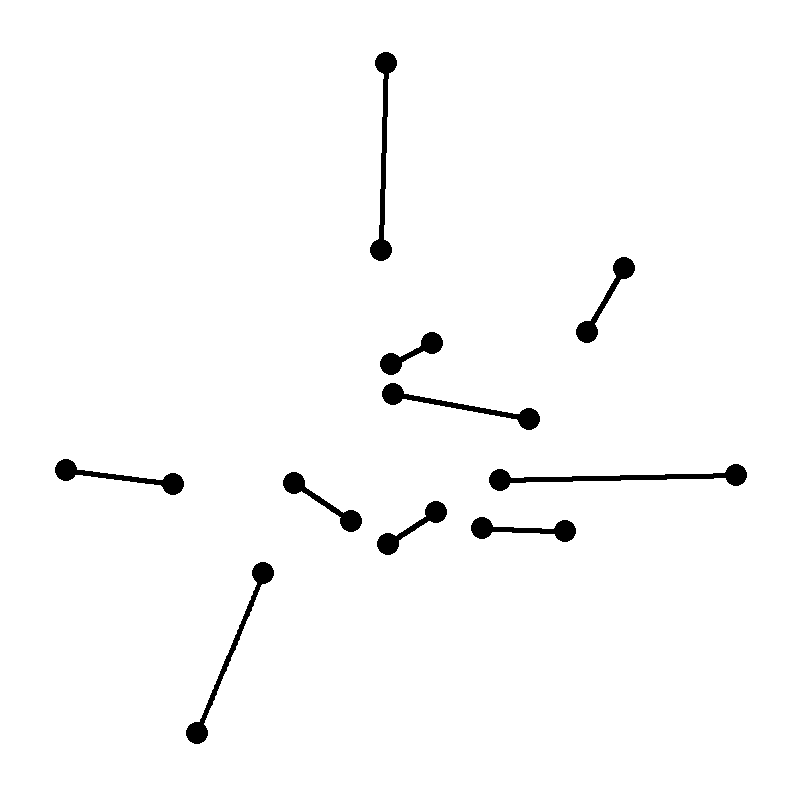}
        \caption{MDP graph}
        \label{fig:graphsB}
    \end{subfigure}
    \begin{subfigure}{0.30\textwidth}
        \includegraphics[width=\textwidth]{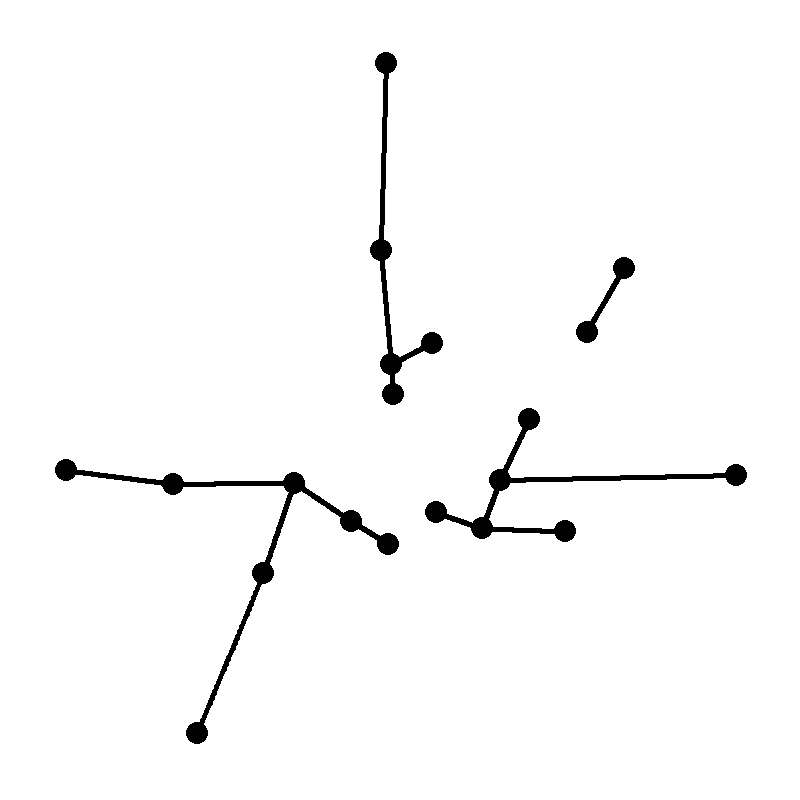}
        \caption{NNL graph}
        \label{fig:graphsC}
    \end{subfigure}
    \caption{{\bf Example graph constructions.} Graphs constructed on $20$ observations which were randomly placed according to a multivariate normal distribution. The graphs are constructed using the following trees: (a) minimal spanning tree (MST), (b) minimal distance pairing (MDP), and (c) nearest neighbour linking (NNL). No orthogonal trees were used.}
    \label{fig:graphs}
\end{figure}

Functional observations $X_i(t),X_j(t)\in L^p[0,1]$ can be related using the appropriate $p$-norm, given in \eqref{eq:lp_norm}. Based on the pairwise distances, edges of the graph can be selected in various ways. We consider three common trees for the construction of the graph. Examples of each tree are given in Figure \ref{fig:graphs} for reference.

Minimal spanning trees (MST) connect all vertices (observations) while minimizing the total distance across the edges \citep{friedman:rafsky:1979}. Under MST, a vertex many have multiple edges and all vertices are connected. An example MST graph is shown in Figure \ref{fig:graphsA}.

Minimal distance pairing (MDP) pairs vertices so that the total distance between all pairs is minimal. If the number of vertices is odd a pseudo-observation with distance $0$ to all vertices is included \citep{rosenbaum:2005}. MDPs connect  vertices such that each vertex has only a single edge and only two vertices are connected to each other. An example MDP graph is shown in Figure \ref{fig:graphsB}.

Nearest neighbour linking (NNL) finds the nearest $N$ neighbors to each vertex, $N\in\{1,\,\ldots,\, n\}$. In NNLs, a vertex has at least $N$ edges and all vertices do not need to be connected together. In this work, we focus on $N=1$. An example NNL graph is shown in Figure \ref{fig:graphsC}.

\begin{figure}
    \centering
    \includegraphics[width=0.4\textwidth]{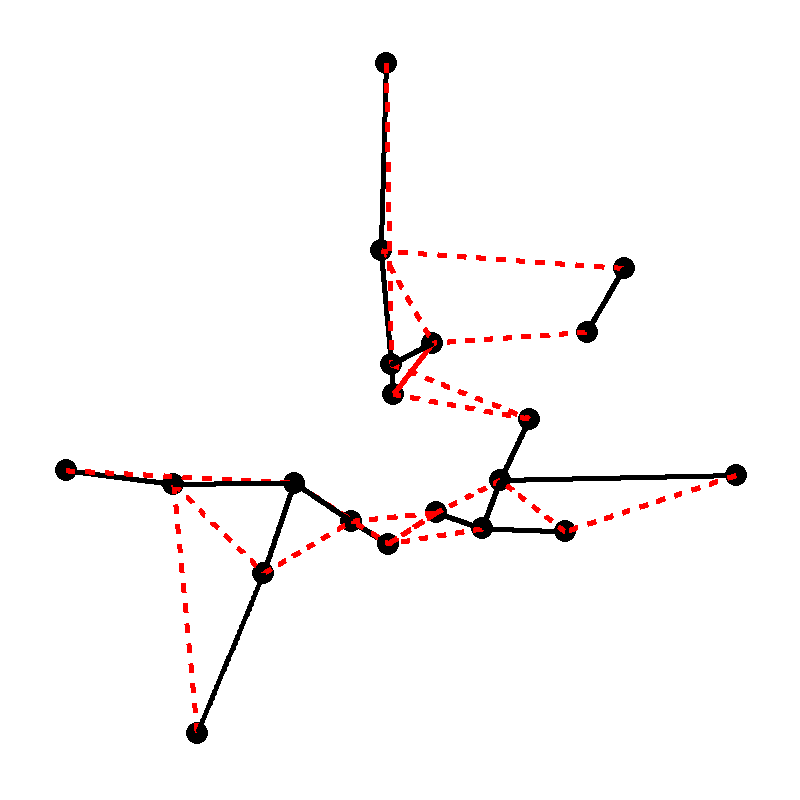}
    \caption{{\bf Example $K$-graph construction.} A NNL-$2$ graph constructed on $20$ observations which were randomly placed according to a multivariate normal distribution. The black lines indicate the first NNL tree; see Figure~\ref{fig:graphsC}. The red lines indicate the second NNL tree, which is orthogonal to the first tree.}
    \label{fig:K_graph}
\end{figure}  

Although not considered in this work, domain-based or weighted trees may also be valuable, e.g. see \cite{nie:nicolae:2021} and \cite{zhou:chen:2022}. Computational considerations may necessitate approximations for graph estimation \citep{liu:chen:2022, chen:chu:2023}.

Intuitively, clusters in the graph may be more clearly defined when graphs are denser, i.e. there are additional edges in the graph. One way to increase the density of a graph is by the union of less dense graphs. A $K$-tree is the union of the first to $K$'th tree. The $i$'th tree is the tree created under the same procedure, e.g. MST, as the previous $1$ to $i-1$ trees subject to the constraint that this tree does not contain any edge that is in the previous trees. Thus, each tree is orthogonal to the other $K-1$ trees. See Figure~\ref{fig:K_graph} for an example of a NNL-$2$. 

The effectiveness of graph-based change point detection when using a $K$-tree is influenced by the choice of $K$. Intuitively, if $K$ is too large or too small then no clusters will be formed regardless of the underlying distribution, and hence there will be no power to detect changes. Section \ref{sec:simulations} investigates the choice of $K$.

\section{Simulations} \label{sec:simulations} 

The practical implications of the proposed tests statistics and related graph-construction parameters are explored in this section. Under various scenarios, several values for the tuning parameters are studied. The graph-based methods are also compared to existing functional methods. When considering changes in the mean, the graph-based methods are compared to the functional mean-change method proposed by \cite{aue:rice:sonmez:2018}. This is denoted {\bf ARS-18}. When considering changes in the covariance kernel, the graph-based methods are compared to the functional method proposed by \cite{horvath:rice:zhao:2022}. This is denoted {\bf HRZ-22}. When considering distributional changes, the graph-based methods are compared to the functional method proposed by \cite{horvath:rice:vanderdoes:2025}. The simulation-based thresholds for the empirical characteristic functional approach is used and the approach is denoted {\bf HRV-25}.

Graph-based change point detection methods have several tuning parameters, such as: tree type, number of orthogonal trees, distance metric, and test statistic. This section investigates the effect of these choices on the effectiveness of graph-based change point detection. We consider MST, MDP, and NNL trees. All trees were constructed via greedy algorithms for computational reasons. We considered $K=1,3,5,7,15$ for the number of orthogonal trees. We focused on $L^1$ and $L^2$ distances. All $4$ graph-based test statistics are considered, and we denote each by their first letter in subsequent tables, i.e. original (O), weighted (W), max-type (M), and generalized (G). 

Simulation studies were conducted looking at the size and power of the methods under various scenarios. In the single change scenarios, changes in the mean, covariance, and distribution were considered. Dependent data were also investigated. In the multiple change scenario, combinations of these changes were considered. Reported simulations had data which were simulated for lengths varying from $n=15$ to $n=250$.

The simulated FTS were generated using an auto-regressive Karhunen-Lo\'eve expansion; see \cite{aue:rice:sonmez:2018}. Denoted {\bf ARKL},  the data were generated such that
\begin{equation} \label{eq:kl}
    X_i(t) =  \delta_i(t) + \sum_{d=1}^4 \xi_{i,d} \phi_{d}(t), \;\; t\in[0,1],
\end{equation}
where $(\phi_1,\dots,\phi_4)$ are typically the first four orthonormal cubic B-spline basis functions, and the vector ${\boldsymbol \xi}_i= (\xi_{i,1},\dots,\xi_{i,4})^\top$ follows a vector autoregression of order one, so that  
\begin{align*}
    {\boldsymbol \xi}_i = \kappa {\boldsymbol \Psi} {\boldsymbol \xi}_{i-1} + \epsilon_i.
\end{align*}
The matrix ${\boldsymbol \Psi}$ is a random $4\times 4$ matrix generated with standard normal entries and then normalized by a constant so that $\|{\boldsymbol \Psi} \|_{F}=1$, where $\| \cdot \|_{F}$ denotes the Frobenius norm. As such $\kappa$ describes the magnitude of serial dependence of the process, which we take as $\kappa=0$ by default to generate independent data. Under the null hypothesis of no change point, we set the mean to be zero ($\delta_i =0$), and the error vector $\epsilon_i$ is assumed to be a mean zero, $4-$variate normal random variable with diagonal  covariance matrix $\mbox{diag}(3,2,1, 0.5)$.

\begin{table}[t]
    \centering
    \tiny 
    
    \caption{{\bf Size table under $L^2$ distance.} Table for the empirical sizes of the graph-based change point detection methods. Combinations of test statistics, tree types, and orthogonal tree counts tuning parameters were examined. Each graph-based test statistic was computed using the $L^2$ distance. Several data lengths were considered, $n=15,25,50,100,200$. In general, the graph-based change point detection methods were well-sized.}

    \begin{tabular}{c  cccc c cccc c cccc }
         \multicolumn{1}{c}{$n=$} & \multicolumn{4}{c}{15} & & \multicolumn{4}{c}{25} & & \multicolumn{4}{c}{50} \\
         \cmidrule{2-15}
         \multicolumn{1}{c}{Statistic =} & \textbf{O} & \textbf{W} & \textbf{M} & \textbf{G} & & \textbf{O} & \textbf{W} & \textbf{M} & \textbf{G} & & \textbf{O} & \textbf{W} & \textbf{M} & \multicolumn{1}{c}{\textbf{G}} \\
        \cmidrule{2 - 15}
        MDP-1  &  0.060 & 0.000 & 0.000 & 0.000 & & 0.040 & 0.031 & 0.031 & 0.033 & & 0.016 & 0.016 & 0.013 & 0.013 \\ 
        MDP-3  &  0.067 & 0.068 & 0.068 & 0.076 & & 0.046 & 0.037 & 0.039 & 0.039 & & 0.060 & 0.060 & 0.054 & 0.055 \\ 
        MDP-5  &  0.070 & 0.069 & 0.069 & 0.073 & & 0.044 & 0.047 & 0.047 & 0.049 & & 0.043 & 0.044 & 0.048 & 0.047 \\ 
        MDP-7  &  0.070 & 0.078 & 0.078 & 0.079 & & 0.057 & 0.051 & 0.051 & 0.047 & & 0.052 & 0.054 & 0.055 & 0.056 \\ 
        MDP-15  &   - &  - &  - &  - & &  - &  - &  - &  - & & 0.055 & 0.055 & 0.057 & 0.057 \\ 
        \cmidrule{2-15}
        NNL-1  &  0.064 & 0.044 & 0.044 & 0.048 & & 0.052 & 0.051 & 0.051 & 0.054 & & 0.061 & 0.059 & 0.059 & 0.052 \\ 
        NNL-3  &  0.069 & 0.059 & 0.059 & 0.057 & & 0.054 & 0.056 & 0.056 & 0.056 & & 0.055 & 0.059 & 0.059 & 0.058 \\ 
        NNL-5  &  0.056 & 0.066 & 0.066 & 0.064 & & 0.056 & 0.054 & 0.054 & 0.056 & & 0.052 & 0.067 & 0.067 & 0.067 \\ 
        NNL-7  &  0.057 & 0.059 & 0.063 & 0.066 & & 0.051 & 0.053 & 0.053 & 0.052 & & 0.048 & 0.065 & 0.065 & 0.060 \\ 
        NNL-15  &  - & - & - & - & & - & - & - & - & & 0.055 & 0.064 & 0.064 & 0.057 \\ 
        \cmidrule{2-15}
        MST-1  &  0.052 & 0.043 & 0.048 & 0.050 & & 0.050 & 0.044 & 0.045 & 0.042 & & 0.062 & 0.058 & 0.057 & 0.063 \\ 
        MST-3  &  0.074 & 0.079 & 0.073 & 0.075 & & 0.054 & 0.049 & 0.047 & 0.042 & & 0.055 & 0.059 & 0.060 & 0.056 \\ 
        MST-5  &  0.063 & 0.065 & 0.066 & 0.071 & & 0.056 & 0.055 & 0.052 & 0.053 & & 0.052 & 0.056 & 0.058 & 0.056 \\ 
        MST-7  &  0.050 & 0.054 & 0.063 & 0.064 & & 0.048 & 0.048 & 0.044 & 0.041 & & 0.041 & 0.059 & 0.063 & 0.057 \\ 
        MST-15  &   - &  - &  - &  - & &  - & - & - &  - & & 0.054 & 0.062 & 0.060 & 0.057 \\ 
        \cmidrule{2-15}
    \end{tabular}
    \vspace{0.5cm}

    \begin{tabular}{c  cccc c cccc }
        \multicolumn{1}{c}{$n=$} & \multicolumn{4}{c}{100} & & \multicolumn{4}{c}{200} \\
        \cmidrule{2-10}
        \multicolumn{1}{c}{Statistic =} & \textbf{O} & \textbf{W} & \textbf{M} & \textbf{G} & & \textbf{O} & \textbf{W} & \textbf{M} & \multicolumn{1}{c}{\textbf{G}} \\
        \cmidrule{2 - 10}
        MDP-1  &  0.063 & 0.063 & 0.061 & 0.065 & & 0.061 & 0.061 & 0.052 & 0.060 \\ 
        MDP-3  &  0.046 & 0.048 & 0.045 & 0.049 & & 0.073 & 0.073 & 0.067 & 0.067 \\ 
        MDP-5  &  0.064 & 0.064 & 0.061 & 0.065 & & 0.054 & 0.054 & 0.055 & 0.052 \\ 
        MDP-7  &  0.059 & 0.060 & 0.057 & 0.053 & & 0.058 & 0.058 & 0.052 & 0.055 \\ 
        MDP-15  &  0.052 & 0.052 & 0.045 & 0.045 & & 0.058 & 0.058 & 0.056 & 0.055 \\ 
         \cmidrule{2 - 10}
        NNL-1  &  0.048 & 0.067 & 0.067 & 0.064 & & 0.060 & 0.054 & 0.053 & 0.060 \\ 
        NNL-3  &  0.047 & 0.054 & 0.053 & 0.052 & & 0.077 & 0.059 & 0.061 & 0.061 \\ 
        NNL-5  &  0.061 & 0.064 & 0.065 & 0.062 & & 0.063 & 0.061 & 0.059 & 0.060 \\ 
        NNL-7  &  0.055 & 0.059 & 0.059 & 0.058 & & 0.066 & 0.055 & 0.055 & 0.051 \\ 
        NNL-15  &  0.045 & 0.054 & 0.054 & 0.053 & & 0.062 & 0.061 & 0.061 & 0.064 \\ 
         \cmidrule{2 - 10}
        MST-1  &  0.047 & 0.044 & 0.047 & 0.054 & & 0.064 & 0.060 & 0.056 & 0.055 \\ 
        MST-3  &  0.049 & 0.045 & 0.044 & 0.052 & & 0.078 & 0.059 & 0.061 & 0.057 \\ 
        MST-5  &  0.055 & 0.058 & 0.058 & 0.051 & & 0.073 & 0.058 & 0.059 & 0.053 \\ 
        MST-7  &  0.065 & 0.058 & 0.059 & 0.061 & & 0.053 & 0.053 & 0.054 & 0.050 \\ 
        MST-15  &  0.051 & 0.055 & 0.055 & 0.054 & & 0.058 & 0.059 & 0.060 & 0.064 \\ 
         \cmidrule{2 - 10}
    \end{tabular}
    \label{tab:null_L2_perm}
\end{table}
Mean changes are created in {\bf ARKL} by setting $\delta_i(t) =\delta \mathds{1}\{ i > k^*\}$, i.e. the mean changes from zero to the function $\delta$ after the change point $k^*$. 

Covariance changes in {\bf ARKL} are created by setting $\delta =0$ and for $i \in \{1,\dots,k^*\}$ letting $\mbox{Cov}(\epsilon_i) = \mbox{diag}(3,2,1, 0.5)$, and for $i \in \{k^*+1,\dots,n\}$, $\mbox{Cov}(\epsilon_i) = \mbox{diag}(3\Delta,2\Delta,\Delta, 0.5\Delta)$. The magnitude of the change in the covariance is characterized by the degree to which $\Delta > 1$.

Distributional changes in {\bf ARKL} are created by first letting $\delta=0$ and $\Sigma = \mbox{diag}(3,2,1, 0.5)$. Then for $i \in \{1,\dots,k^*\}$ we generate $\epsilon_i \stackrel{iid}{\sim} \mathcal{N}(0, \Sigma)$, where $\mathcal{N}(0, \Sigma)$ denotes a multivariate normal distribution with mean zero and covariance matrix $\Sigma$, and for $i \in \{k^*+1,\dots,n\}$ we generate $\epsilon_i = (\epsilon_{i,1},...,\epsilon_{i,5})^\top$ so that $\epsilon_{i,j} \stackrel{iid}{\sim} (1/j^{1/4})\Gamma_s(\eta, 1)$, where $\Gamma_s(\eta, 1)$ denotes a gamma distribution with scale parameter $1$ and shape parameter $\eta$ that has been standardized to have mean $0$ and variance $1$. In this case the mean function and covariance operator of the sample is homogeneous, but the distribution changes at $k^*$. Note that $\eta$ tending to infinity causes $H_0$ to hold, so the magnitude of the change is larger for smaller values of $\eta$. We also consider an centered exponential distribution created under the same framework.

For brevity, when the data were generated via {\bf ARKL} only the parameters that were different from the previously stated defaults are given. Additionally, $1,000$ independent simulations per scenario were conducted unless otherwise stated. In order to detect changes, the random shuffling method was used to compute the $p$-values, $M=1,000$. Reported $p$-values were compared to the nominal significance level of $\alpha=0.05$.

\subsection{No change simulations} \label{sec:sim_null}

\begin{table}[t]
    \centering
    \tiny 
    \caption{{\bf Size table under varying distance metrics.} Table for the empirical sizes of the graph-based change point detection statistics with varying tree types and orthogonal tree counts using $L^2$ and $L^1$ distance metrics. Data were length $n=50$. Both $L^2$ and $L^1$ were typically well-sized, with the exception of MDP-$1$ which was very conservative.}
    \begin{tabular}{c  cccc c cccc  }
        \multicolumn{1}{c}{} & \multicolumn{9}{c}{\textbf{Original Data}} \\
        \multicolumn{1}{c}{Distance =} & \multicolumn{4}{c}{\textbf{$L^1$}} & & \multicolumn{4}{c}{\textbf{$L^2$}} \\
        \cmidrule{2-10}
        \multicolumn{1}{c}{Statistic =} & \textbf{O} & \textbf{W} & \textbf{M} & \textbf{G} & & \textbf{O} & \textbf{W} & \textbf{M} & \multicolumn{1}{c}{\textbf{G}} \\
        \cmidrule{2 - 10}
        MDP-1 & 0.018 & 0.018 & 0.018 & 0.018 & & 0.010 & 0.010 & 0.010 & 0.010 \\ 
        MDP-3 & 0.063 & 0.063 & 0.057 & 0.061 & & 0.051 & 0.051 & 0.046 & 0.048 \\ 
        MDP-5 & 0.054 & 0.054 & 0.057 & 0.057 & & 0.057 & 0.057 & 0.056 & 0.056 \\ 
        MDP-7 & 0.056 & 0.056 & 0.057 & 0.056 & & 0.053 & 0.055 & 0.061 & 0.059 \\ 
        MDP-15 & 0.052 & 0.052 & 0.055 & 0.055 & & 0.056 & 0.057 & 0.054 & 0.054 \\ 
        \cmidrule{2 - 10}
        NNL-1 & 0.056 & 0.059 & 0.060 & 0.062 & & 0.050 & 0.052 & 0.052 & 0.049 \\
        NNL3 & 0.052 & 0.051 & 0.051 & 0.05 & & 0.059 & 0.063 & 0.064 & 0.069 \\ 
        NNL5 & 0.061 & 0.052 & 0.051 & 0.046 & & 0.058 & 0.064 & 0.064 & 0.067 \\ 
        NNL7 & 0.061 & 0.052 & 0.052 & 0.058 & & 0.058 & 0.057 & 0.057 & 0.060 \\ 
        NNL15 & 0.061 & 0.051 & 0.051 & 0.052 & & 0.053 & 0.057 & 0.057 & 0.060 \\ 
        \cmidrule{2 - 10}
        MST-1 & 0.058 & 0.064 & 0.064 & 0.064 & & 0.046 & 0.059 & 0.056 & 0.064 \\ 
        MST-3 & 0.056 & 0.054 & 0.053 & 0.046 & & 0.06 & 0.068 & 0.067 & 0.061 \\ 
        MST-5 & 0.065 & 0.056 & 0.055 & 0.060 & & 0.056 & 0.062 & 0.064 & 0.062 \\ 
        MST-7 & 0.057 & 0.055 & 0.058 & 0.059 & & 0.065 & 0.053 & 0.053 & 0.051 \\ 
        MST-15 & 0.052 & 0.046 & 0.049 & 0.054 & & 0.051 & 0.047 & 0.048 & 0.056 \\ 
        \cmidrule{2 - 10}
    \end{tabular}
    \label{tab:distance}
\end{table}

Under the no-change null hypothesis, Table \ref{tab:null_L2_perm} investigates the empirical size of the graph-based tests using the $L^2$ distance metric. Table \ref{tab:null_L2_perm} shows that for all investigated sample sizes--$n=15, 25, 50, 100, 200$--the graph-based tests were reasonably well-sized. The various edge-count metrics and tree types performed similarly. An increasing number of orthogonal trees resulted in similar or better sizes, provided the sample was sufficiently long. Since change point detection is often considered more difficult for smaller samples, we note that when $n=50$ the graph-based methods are well-sized. Throughout this section, we use $n=50$ for simulations.

Table \ref{tab:distance} examines the size of the test statistic under $L^1$ and $L^2$ distance metrics. Both $L^1$ and $L^2$ perform well over the various scenarios. Unreported values of $L^p$ generally performed well, with the exception that $L^{\infty}$ was generally poorly-sized. We report the remaining simulations in this section using the $L^2$ distance, although similar results were seen for $L^1$.

\begin{figure}[thp]
    \centering
    \begin{subfigure}{0.45\textwidth}
        \includegraphics[width=0.825\textwidth]{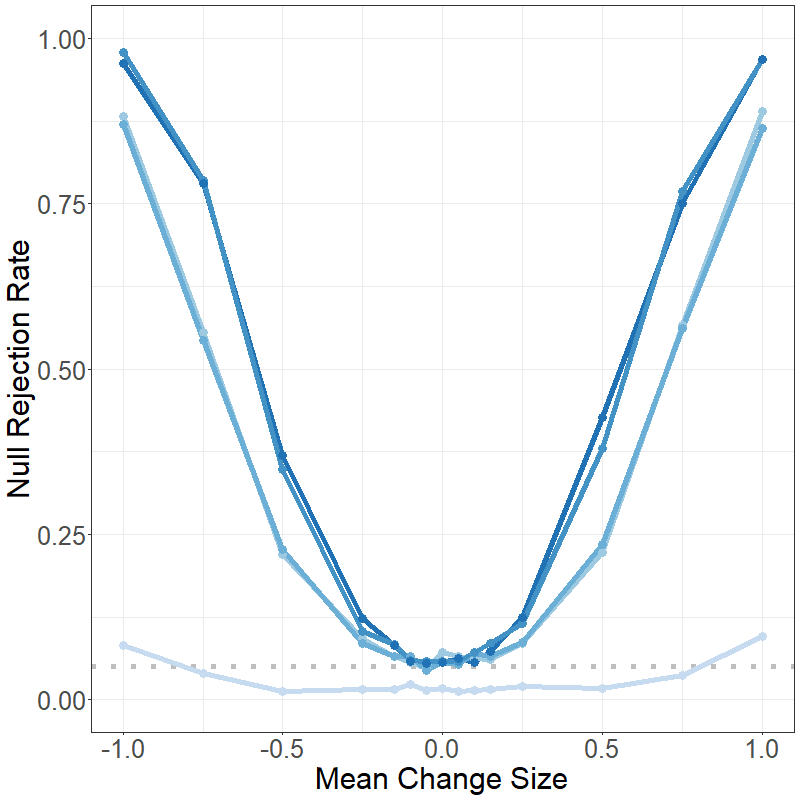}
        \caption{MDP graph}
        \label{fig:numTrees1}
    \end{subfigure}
    \begin{subfigure}{0.45\textwidth}
        \includegraphics[width=0.825\textwidth]{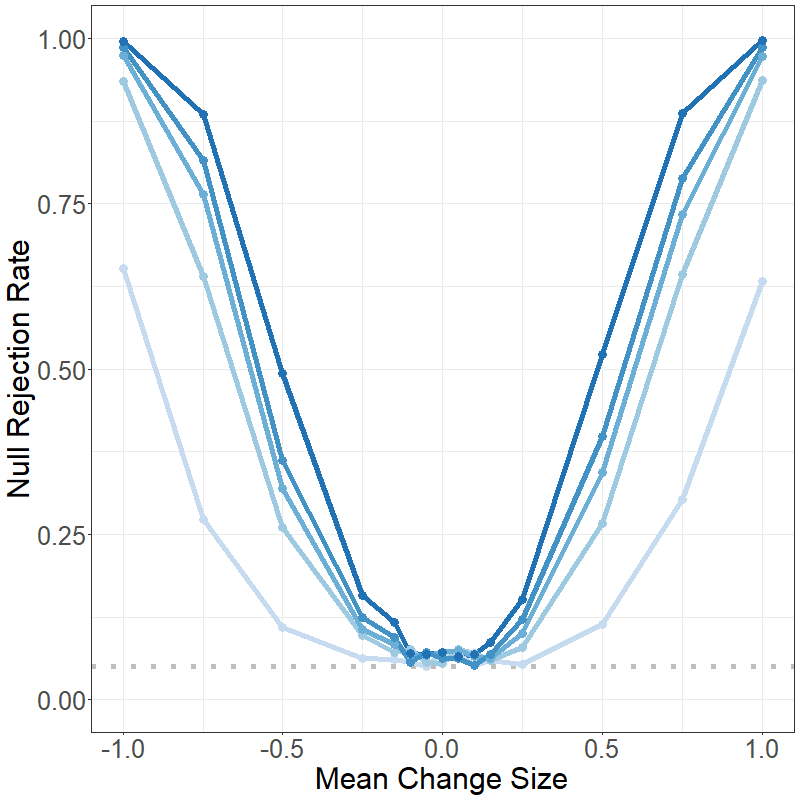}
        \caption{MST graph}
        \label{fig:numTrees2}
    \end{subfigure}
    \\
    \begin{subfigure}{0.45\textwidth}
        \includegraphics[width=0.825\textwidth]{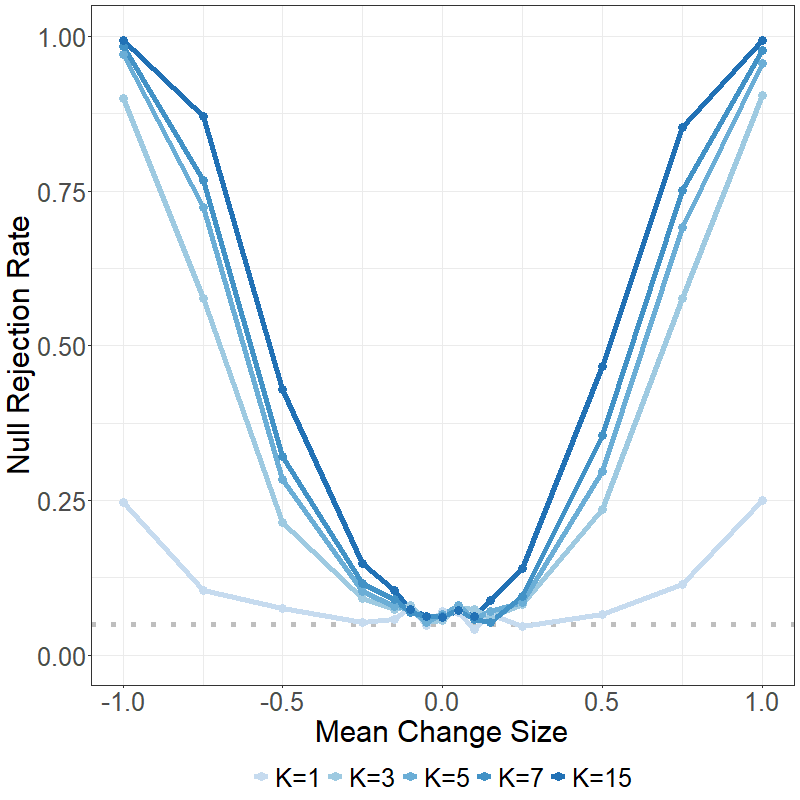}
        \caption{NNL graph}
        \label{fig:numTrees3}
    \end{subfigure}
    \caption{{\bf Mean power curves for normal errors.} Power curves as a function of the change magnitude based on mid-sample mean changes. The change magnitude $\delta$ varied from $-1$ to $1$ with samples of length $n=50$. The number of orthogonal trees, $K$, are colored by increasing blue intensity. The dotted horizontal gray line is the nominal significance level. Power curves were constructed using the generalized test statistic as in \eqref{eq:teststat_g}. Power curves by graph types are displayed in each figure: (a)~MDP, (b)~MST, and (c)~NNL. Power tended to increase with a larger number of orthogonal trees. MST and NNL were marginally more powerful than MDP.}
    \label{fig:numTrees}
\end{figure}

\subsection{Single change simulations} \label{sec:sim_single}

\begin{figure}[t]
    \centering
    \includegraphics[width=0.95\textwidth]{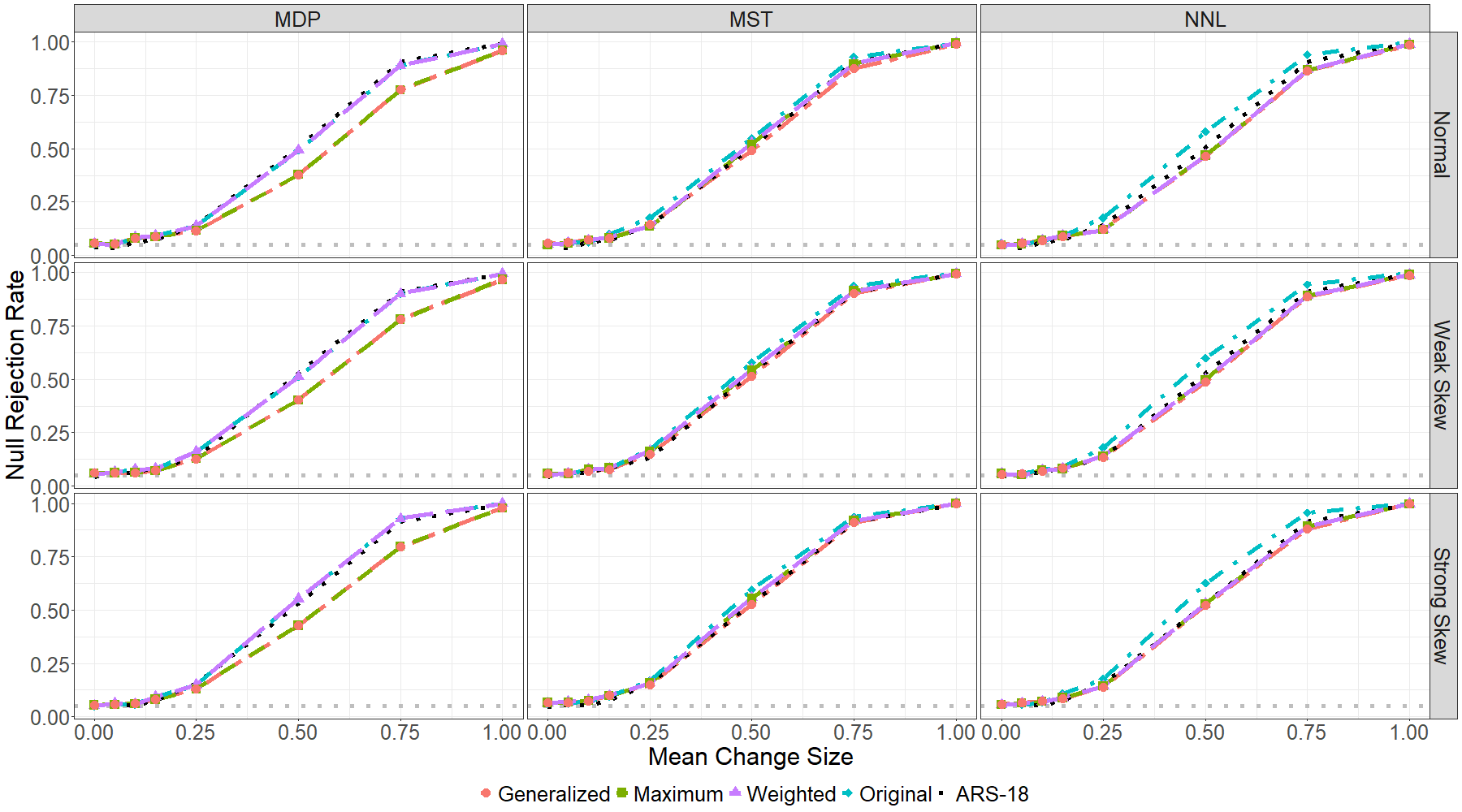}
    \caption{{\bf Mean power curves for skewed errors.} Power curves as a function of the change magnitude based on mid-sample mean changes with normal and skewed-normal error distributions. The change magnitude $\delta$ varied from $0$ to $1$ with samples of length $n=50$. The top plot used data normal errors while the middle and bottom plots used increasingly skewed-normal errors. The methods are generally comparable. The weighted and original methods performed better than the generalized or max-type statistic, especially for the MDP graph. The original test statistic exhibited power comparable to {\bf ARS-18}. }
    \label{fig:normal_skew}
\end{figure}

Under the AMOC alternative, we considered the effect of the number of orthogonal trees. Figure \ref{fig:numTrees} examines the effect of the number of orthogonal trees, $K$, for mean change point detection. The data were simulated using {\bf ARKL} as in \eqref{eq:kl} for length $n=50$ samples. The change magnitudes $\delta=0,\pm0.05, \pm0.1, \pm0.15, \pm0.25, \pm0.5, \pm0.75, \pm1$ were considered for the mid-sample change. The errors were normally distributed. The results when using the generalized test statistic are shown in \eqref{eq:teststat_g}; unreported simulations showed similar patterns for the other graph-based test statistics.

For each graph construction, increasing the number of orthogonal trees tended to increase the power of the test. However, the increased power comes at the cost of requiring additional data and computational time to fit the graph and assuming that the changes are sufficiently separated. Hence, while increasing $K$ can lead to increased power, it may overlook changes. Thus, the choice of $K$ is strongly data dependent.

\begin{figure}[t]
    \centering
    \includegraphics[width=0.95\textwidth]{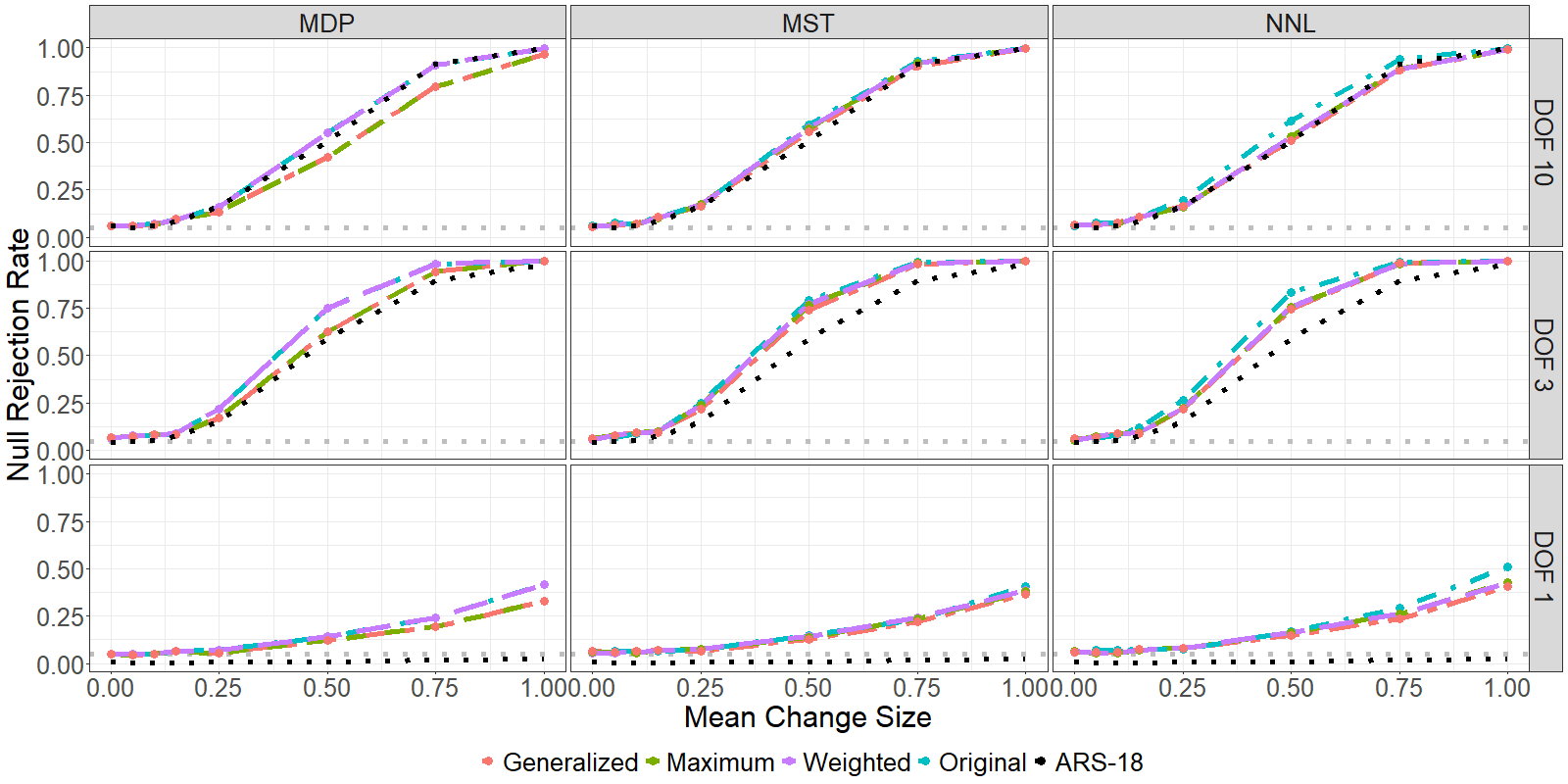}
    \caption{{\bf Mean power curves for $t$-distribution errors.} Power curves as a function of the change magnitude for a mid-sample mean change with errors distributed according to a $t$-distribution. The change magnitude $\delta$ varied from $0$ to $1$ with samples of length $n=50$. The dotted horizontal gray line is the nominal significance level. The top figure was generated from data with $t_{10}$ errors, the middle with $t_3$ errors, and the bottom with $t_{1}$ errors; see also Figure \ref{fig:normal_skew} for power under Gaussian errors. As the normality assumption was increasingly violated, the graph-based methods increasingly outperformed {\bf ARS-18} in terms of size and power. The graph types performed similarly. The test statistics were also similar, although original and weighted typically performed better than generalized and max-type. }
    \label{fig:t_dist}
\end{figure}

In this section, we consider only $15$-trees due to the increased power and its additional robustness to non-normal errors. Nonetheless, typically all choices are reasonable, with the notable exception of $1$-trees often having low power. Yet even the power loss of $1$-trees can be mitigated using max-type or generalized test statistics. Therefore, the conclusions in this section generally apply to any number of orthogonal trees. When the distance between changes is unknown, a smaller value for $K$ is recommended. Practically it may be valuable to begin with a larger $K$ and shrink $K$ when no changes are detected to ensure no edge changes are present.

Power curves investigating the single mean change alternative under normal and normally-skewed errors are given in Figure \ref{fig:normal_skew} for $n=50$. In each simulation, the first half of the sample had mean $0$ which abruptly changed at the midpoint to the second half mean, $\delta=0, 0.05, 0.1, 0.15, 0.25, 0.5, 0.75, 1$. The graph-based methods exhibited similar power to each other and {\bf ARS-18}. As expected, the weighted and original methods, which are tuned for mean changes, performed better than the other graph-based methods in all cases, although the differences were typically small.

\begin{figure}[t]
    \centering
    \includegraphics[width=0.95\textwidth]{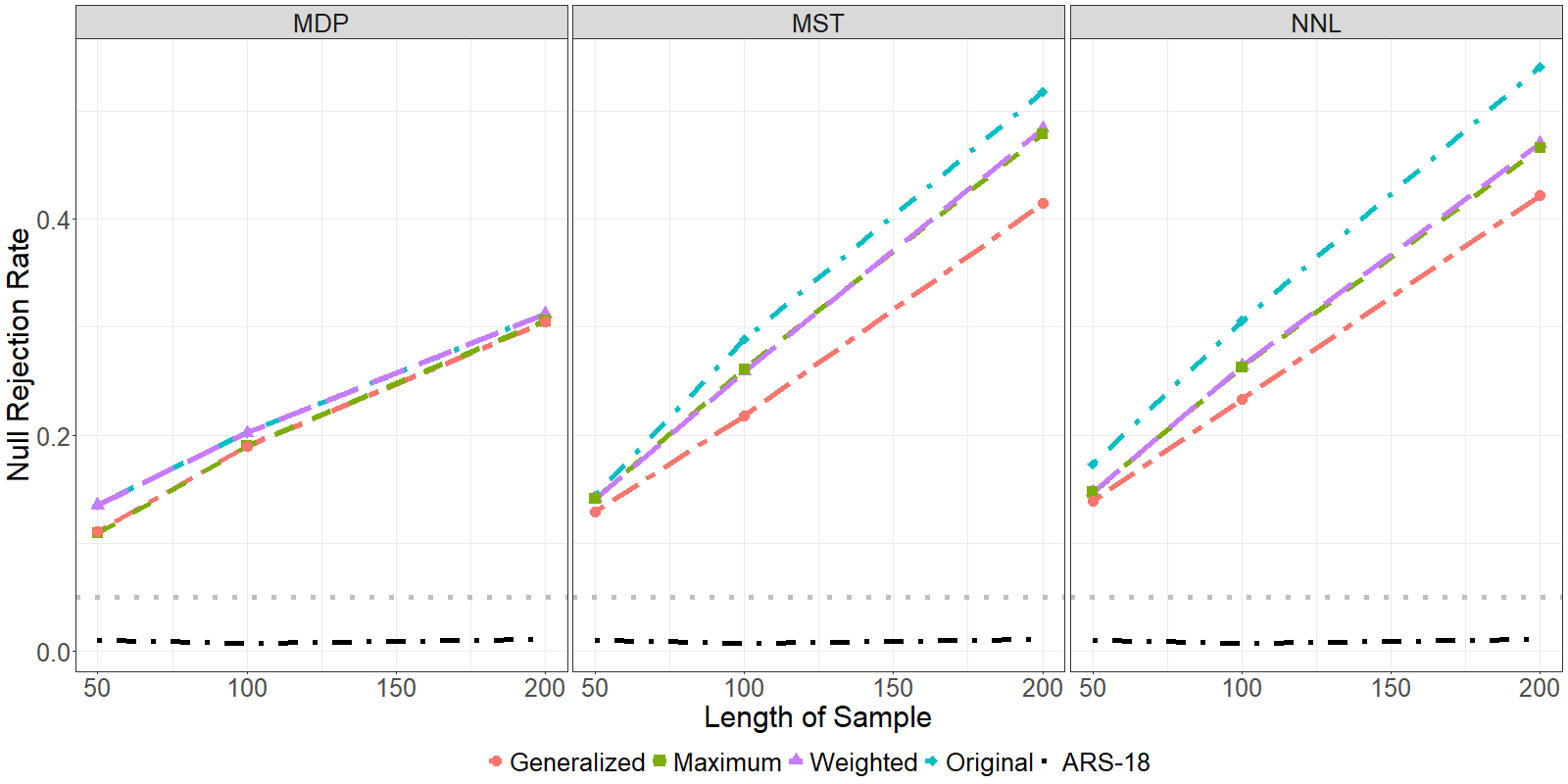}
    \caption{{\bf Power curves for $t_1$ errors with varying data length.} Power curves as a function of data length for a mid-sample mean change with $t_1$ errors. The change magnitude $\delta$ was set to $0.5$ for data lengths of $n=50,100,200$. The dotted horizontal gray line is the nominal significance level.  Improvement over increasing data length $n$ is observed for all graph-based approaches, but not for {\bf ARS-18}. MST and NNL performed better than MDP. The original statistic performed better than the weighted and max-type. The weighted and max-type statistics performed better than the generalized statistic.}
    \label{fig:len_t1}
\end{figure}

Figure \ref{fig:t_dist} compares the power of graph-based approaches and {\bf ARS-18} for mean changes under varying levels of non-normality. Mean changes were considered such that $\delta=0, 0.05, 0.1, 0.15, 0.25, 0.5, 0.75. 1$. The non-normality came in the form of heavy tails with the errors distributions being $t_{10}$, $t_3$, and $t_1$. Figure \ref{fig:t_dist} shows that the graph-based methods were more robust and outperformed {\bf ARS-18} when the error distributions had heavier tails. Again the weighted and original methods, which are tuned to the mean change, performed better than the generalized and max-type graph-based methods.

A clear loss in power was observed as the errors became increasingly heavy-tailed; see Figure \ref{fig:t_dist}. The effects of data length, $n$, on this power loss are considered in Figure \ref{fig:len_t1}. Figure \ref{fig:len_t1} shows the power of the graph-based and {\bf ARS-18} methods for $n=50,100,200$ when the errors were distributed according to a $t_1$ distribution. The data were generated using the {\bf ARKL} as in \eqref{eq:kl} with a small, centered mean change of $\delta=0.5$. The {\bf ARS-18} did not benefit from increased sample sizes while the graph-based approaches increased in power. Interestingly, MDP does not benefit as heavily from increasing $n$ when compared to MST and NNL. The original statistic performed better than the weighted and max-type statistics. The weighted and max-type statistics performed better than the generalized statistic.

\begin{figure}[th]
    \centering
    \includegraphics[width=0.95\textwidth]{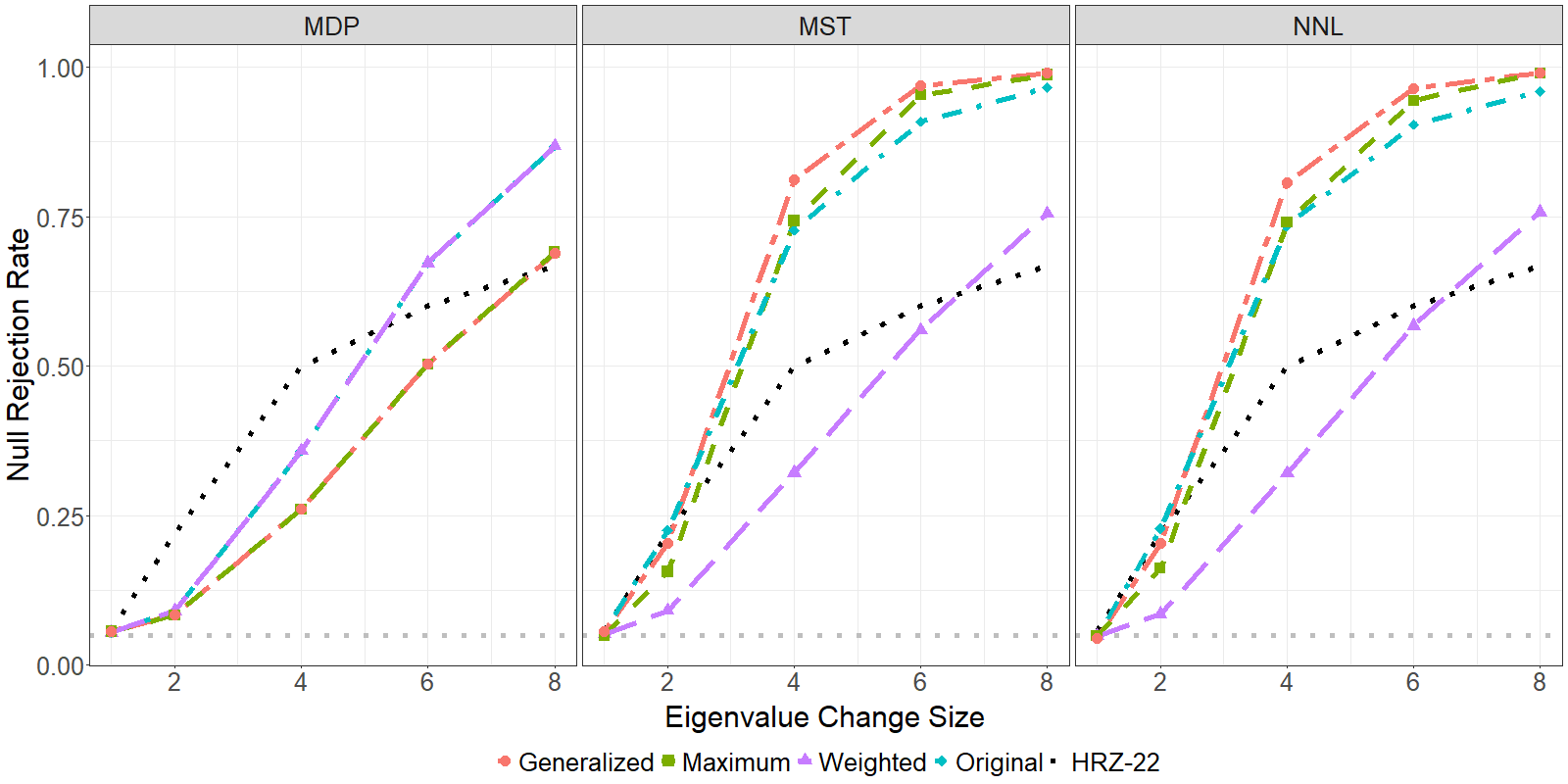}
    \caption{{\bf Eigenvalue change power curves.} Power curves as a function of the change magnitude based on mid-sample eigenvalue changes. The change magnitude $\Delta$ was varied from $1$ to $8$ for sample length $n=50$. The dotted horizontal gray line is the nominal significance level. For MST and NNL graphs, the graph-based statistics quickly surpassed {\bf HRZ-22} in terms of power. MDP graphs performed worse than {\bf HRZ-22} for smaller eigenvalue changes and better for larger eigenvalue changes. Interestingly, the weighted statistic was not strongly influenced by the graph type, where the other graph-based test statistics exhibited increased power for MST and NNL graphs.}
    \label{fig:eig_change}
\end{figure}

A valuable feature of the graph-based approaches is their ability to capture changes of various nature, e.g. changes in the data mean or changes in the error distributions. Detecting changes in the eigenvalues of the models innovation covariance is shown in Figure \ref{fig:eig_change}. For $n=50$, data were generated per {\bf ARKL} as in \eqref{eq:kl} with normal errors. The eigenvalue changes were considered for $\Delta=1, 2, 4, 6, 8$. {\bf HRZ-22} outperformed the graph-based approaches for small changes, but was quickly surpassed as the change magnitude became more extreme. 

\begin{figure}[t]
    \centering
    \includegraphics[width=0.95\textwidth]{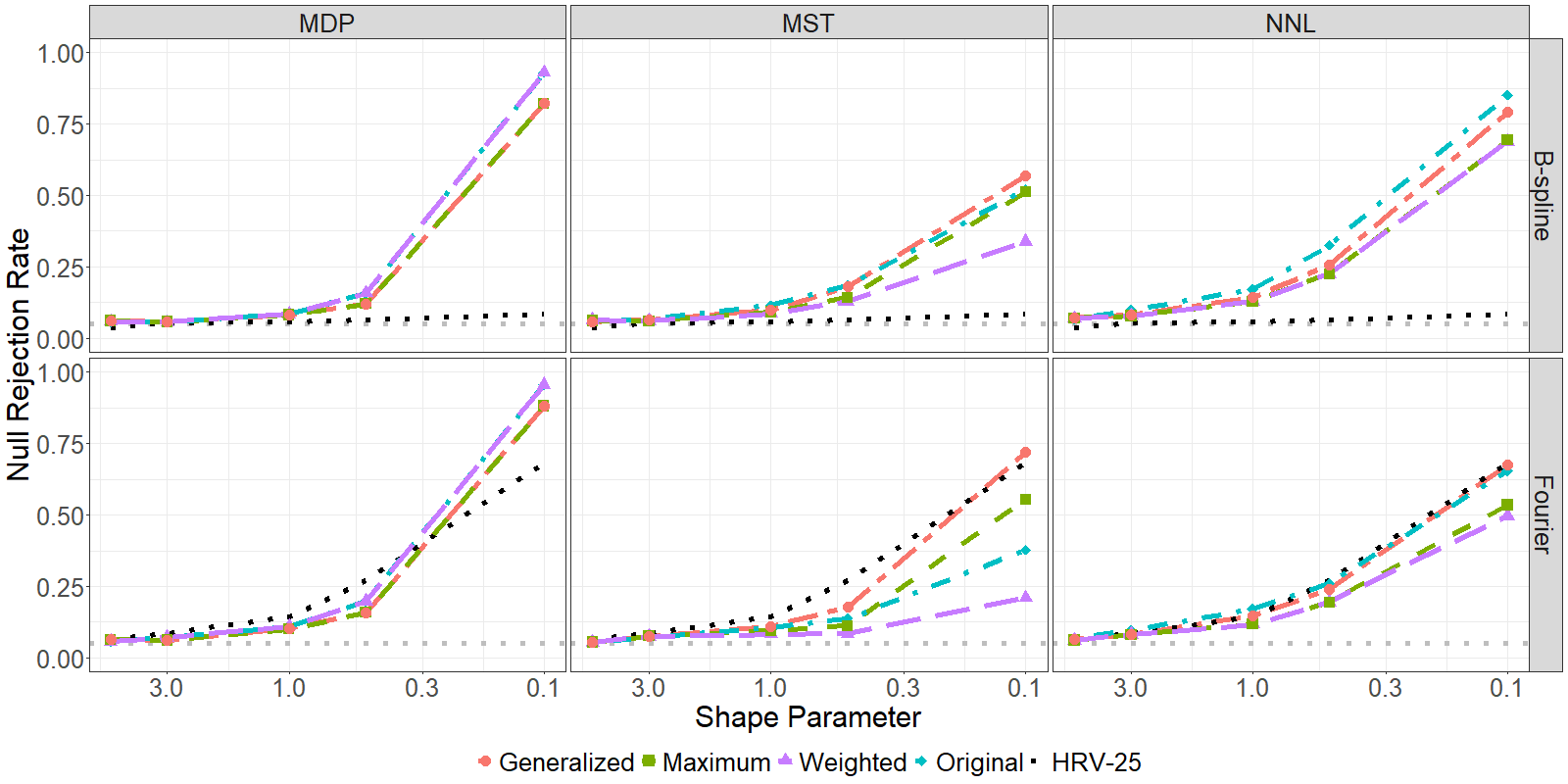}
    \caption{{\bf Distributional change power curves.} Power curves as a function of the change magnitude based on mid-sample distributional changes. The change magnitude $\eta$ was varied from $5$ to $0.1$ for sample length $n=50$.  The dotted horizontal gray line is the nominal significance level. Both the graph-based and {\bf HRV-25} methods improved with increased  distributional differences. The graph-based methods performed better for extreme cases and for the B-spline basis, but {\bf HRV-25} performed better for moderate changes with the Fourier basis.}   
    \label{fig:distributionalChanges}
\end{figure}

Figure \ref{fig:distributionalChanges} examines the power of the methods when detecting distributional changes midway through data of length $n=50$. The first half of the data had errors with a normal distribution, and the second half had errors with a standardized Gamma distribution. The shape parameter values considered were $\eta=5,3,1,0.5,0.1$. The graph-based and {\bf HRV-25} methods both exhibited power in detection of distributional changes. The power increased for more significant changes. The graph-based methods performed better than {\bf HRV-25} for large differences and when using the B-spline basis. {\bf HRV-25} performed better for moderate changes with the Fourier basis. Consideration of the data structure is crucial when selecting the optimal approach for detection of distributional changes.

\begin{figure}[t]
    \centering
    \includegraphics[width=0.95\textwidth]{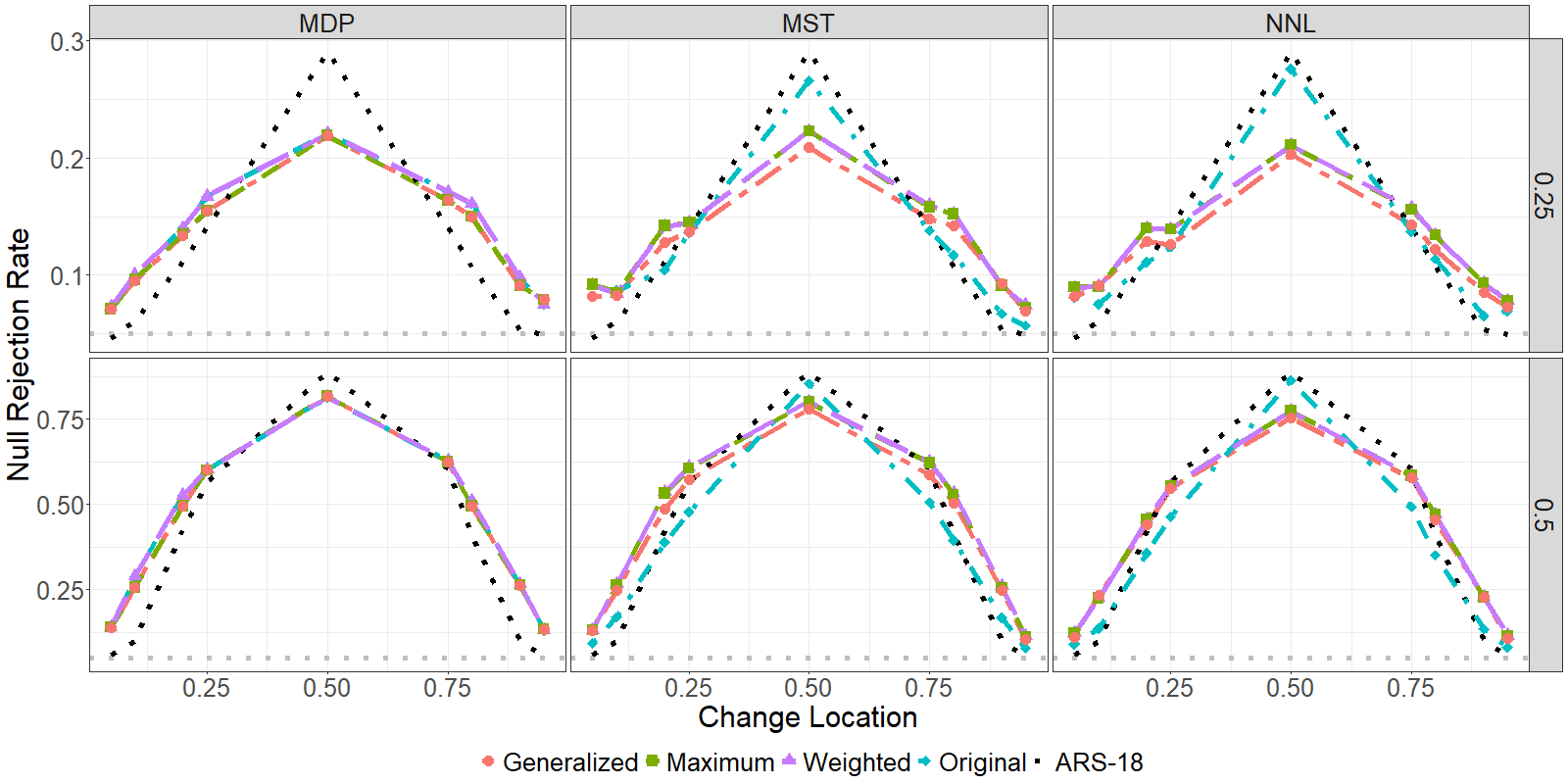}
    \caption{{\bf Mean change location power curves.} Power curves  as a function of the change location based on mean changes. Change magnitudes of $\delta=0.25,0.5$ for length $n=100$ data was considered. The change location $k^*$ was varied from $5$ to $95$. The dotted horizontal gray line is the nominal significance level. The graph-based approaches exhibited more robustness to changes at the edges of the data than {\bf ARS-18}. {\bf ARS-18} exhibited more power for mid-sample changes, particularly for mild changes. The weighted statistic performed best among graph-based methods, particularly with MST and NNL graph types.}
    \label{fig:location_power}
\end{figure}

Often real data have changes skewed to one end of the sample. Thus, we also considered the performance of the models when the location of the change was modified. For $n=100$, change point locations of $k=5, 10, 20, 25, 50, 75, 80, 90, 95$ were examined. At each change location, mild ($\delta=0.25$) and moderate ($\delta=0.5$) mean changes were considered. Figure \ref{fig:location_power} shows the graph-based methods were more robust than {\bf ARS-18} to changes at the edge of the sample. Yet, {\bf ARS-18} exhibited better detection under mid-sample mean changes, specifically when the change was mild.

\begin{figure}[t]
    \centering
    \includegraphics[width=0.95\textwidth]{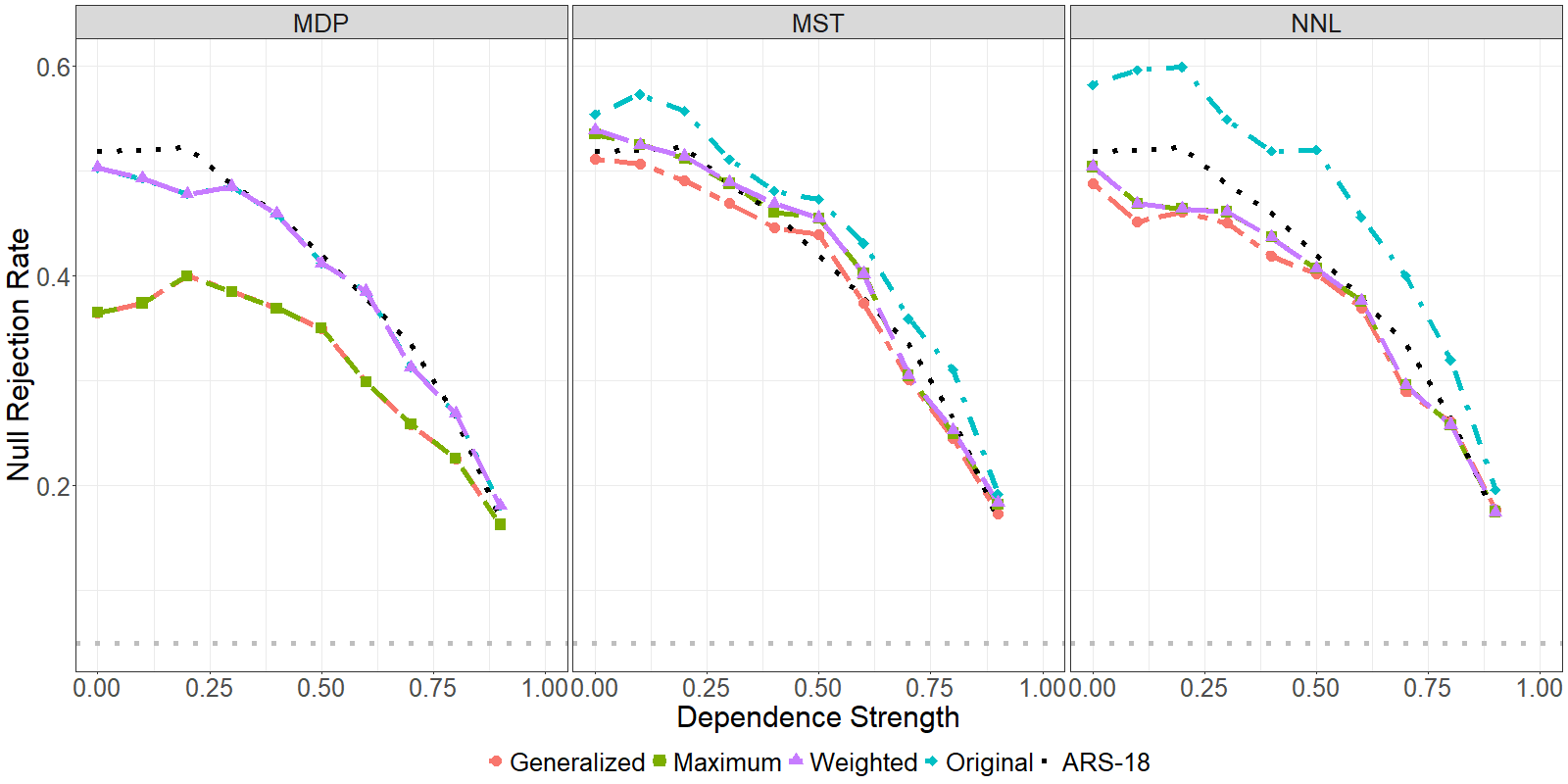}
    \caption{{\bf Size-adjusted mean change power curves under dependency.} Power curves as a function of dependency magnitude based on mid-sample mean changes. Change magnitude $\delta=0.5$ for $n=50$ length data was considered. The dependency magnitude $\kappa$ was varied from $0$ to $0.9$. The size-adjusted rejection rate is reported. Loss of power due to dependency is typically comparable between graph-based and {\bf ARS-18} methods, while MST and NNL performed better than MDP.}
    \label{fig:dependenceMean}
\end{figure}

The graph-based change point detection approach proposed in section \ref{sec:model} is based on serial independence of the data. When the data are serially dependent, models can often be used to capture the dependence, and then change point analysis conducted on the independent, or near-independent, residuals; see section \ref{sec:graph_electricity}. Yet sometimes dependency exists undetected in the data or remains in model residuals. Hence it is valuable to consider the robustness of the graph-based methods under dependency. Figure \ref{fig:dependenceMean} examines the ability of the method to detect a mean change in data with dependence, $\kappa=0, 0.1, \dots, 0.9$. As expected, power decreased as the strength of the dependency increased when accounting for inflated size. The graph-based and {\bf ARS-18} methods exhibited similar power degradation over the increased dependence. The choice of graph played a significant role in the power, with MST and NNL more robust than MDP. We note that while we assumed independence, {\bf ARS-18} can be modified such that estimation is more robust under dependence.

In terms of a single change point, the graph-based approaches were well sized and exhibited good power under a variety of scenarios. The performance of graph-based change point detection methods were generally similar or better than existing functional change point detection methods. Robustness to tuning parameter selections and assumption violations was observed.

\subsection{Multiple change simulations} \label{sec:sim_mult}

\begin{figure}[thp]
    \centering
    \includegraphics[width=0.85\textwidth]{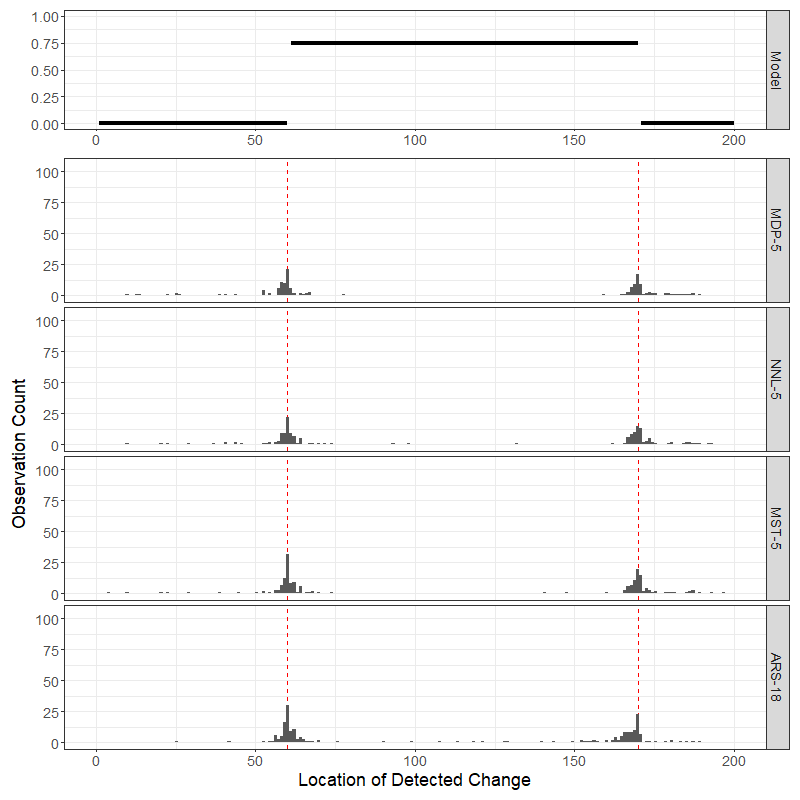}
    \caption{{\bf Multiple mean change location estimates.} Estimates of mean change point locations based on data with $2$ changes. For data of length $n=200$,  the magnitude change $\delta$ changed from $0$ to $0.75$ to $0$ at locations $k^*_1=60$ and $k^*_2=170$. The true mean model is given in the top plot. Histograms for the estimated change point locations from $100$ simulations are given in the lower plots. Vertical red lines indicate the locations of the true changes. The graph-based methods and {\bf ARS-18} show robustness to the changed interval, or epidemic change, scenario. The graph-based methods had more early sample false detections while {\bf ARS-18} had more mid-sample false detections. The graph-based methods were generally comparable. }
    \label{fig:multipleChanges_mean}
\end{figure}

While single change simulations are valuable to understand performance of a method due to a specific change, often multiple change simulations better reflect reality. We applied binary segmentation (BS) to detect multiple changes for the graph-based and existing functional change point detection methods. Using only BS highlighted the differences between approaches; however, other multiple change point extensions may improve the power for all approaches. In this section $K=5$ for orthogonal trees to mimic real-life scenarios when the structure of the data is largely unknown. Similar results were observed when $K=15$. The max-type test statistic in \eqref{test-def:4} is used for the graph-based methods due to its robustness in the single change scenarios. In each setting, only $100$ simulations were run as this was sufficient to observe the detection trends.

Detected change locations for the graph-based and {\bf ARS-18} methods were recorded in the mean changed interval, or epidemic, scenario. Data of length $n=200$ were simulated with the mean $\delta$  changing from $0$ to $0.75$ to $0$. The changes occurred at $60$ and $170$. $100$ simulations were run. The true mean function and histograms of the detected locations are given in Figure \ref{fig:multipleChanges_mean}. The graph-based and {\bf ARS-18} methods typically detected changes in the neighborhood of the true changes. However, the graph-based methods had more early sample false detections and {\bf ARS-18} had more mid-sample false detections.

\begin{figure}[thp]
    \centering
    \includegraphics[width=0.85\textwidth]{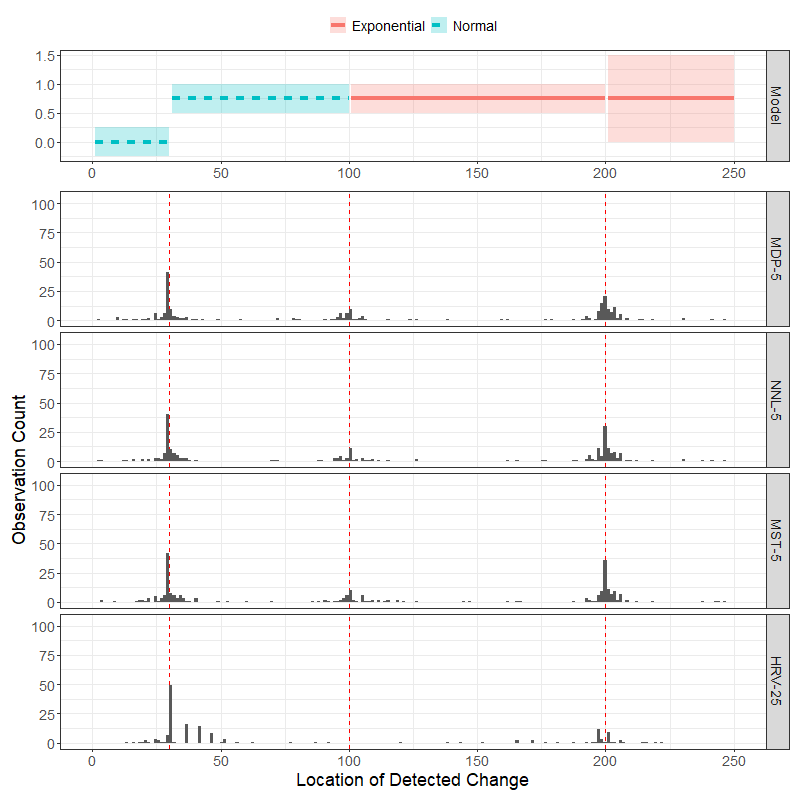}
    \caption{{\bf Multiple distributional change location estimates.} Estimates of distributional change point locations based on data with $3$ changes. For data of length $n=250$, the data changed at locations $k^*_1=30$, $k^*_2=100$, and $k^*_3=200$. A mean change $\delta=0.75$ occurred at $k^*_1$. A distributional change for the errors $\epsilon$'s changed from normal to exponential, both standardized, at $k^*_2$. A covariance change $\Delta=3$ occurred at $k^*_3$. A representation of the true model is given in the top plot. Histograms for the estimated change point locations from $100$ simulations are given in the lower plots.  Vertical red lines indicate the locations of the true changes. The graph-based methods and {\bf HRV-25} commonly detected the first change, with {\bf HRV-25} perhaps performing slightly better. However, the graph-based methods detected the distributional and variance changes far more commonly.
    }
    \label{fig:multipleChanges}
\end{figure}

Figure \ref{fig:multipleChanges} considers $100$ simulations of $n=250$ length samples containing $3$ changes of different types. The graph-based and {\bf HRV-25} methods exhibited power on the first mean change. However, the graph-based methods exhibited more power for the other changes. Per Figure \ref{fig:distributionalChanges}, we suppose the power of the {\bf HRV-25} method would improve using a different basis in {\bf ARKL}, e.g. a Fourier basis. The robustness of parameter choices is of note for the graph-based methods.

\section{Applications} \label{sec:applications} 

This section considers three real-world examples, related to pedestrian counts, stock returns, and electricity prices. For brevity, and due to its strong performance in section \ref{sec:simulations}, we only report the results from the max-type test statistic defined in \eqref{test-def:4} based on MST-$15$ using $L^2$ distance. Nominal significance of $\alpha=0.05$ is used. Multiple changes were detected using binary segmentation (BS). An unreported sensitivity analysis on the parameter choices showed robustness to these choices.

\subsection{Australian pedestrian counts}

\begin{figure}[t]
    \centering
    \includegraphics[trim=7cm 1cm 2.5cm 5.5cm,width=0.6\textwidth,clip]{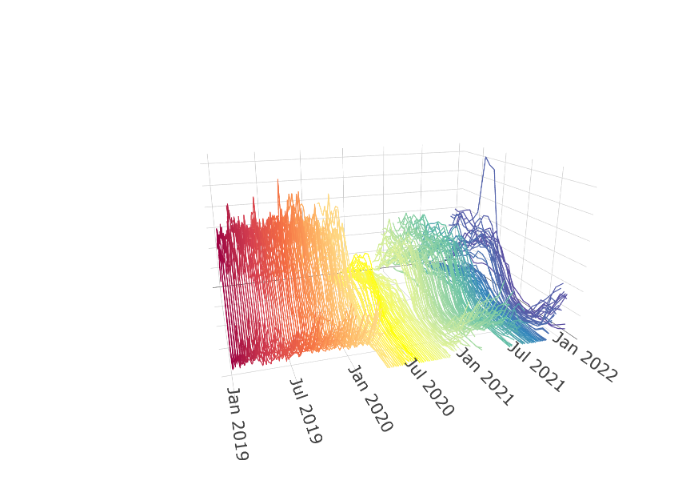}
    \caption{{\bf Melbourne pedestrian counts.} Hourly pedestrian counts on a street in Melbourne Australia from 2019 to 2021.}
    \label{fig:Australia_data}
\end{figure}

The city of Melbourne Australia uses a system of automated sensors to monitor information about pedestrian activity throughout the city. These observations are used to inform development and promote city vibrancy \citep{Melbourne}. We are interested in potential changes in pedestrian count data collected at one of the monitored crossroads from 1 January 2019 to 31 December 2021. The data were recorded every hour and are shown in Figure \ref{fig:Australia_data}.

\begin{figure}[th]
    \centering
    \includegraphics[trim=7cm 1cm 2.5cm 5.5cm,width=0.6\textwidth,clip]{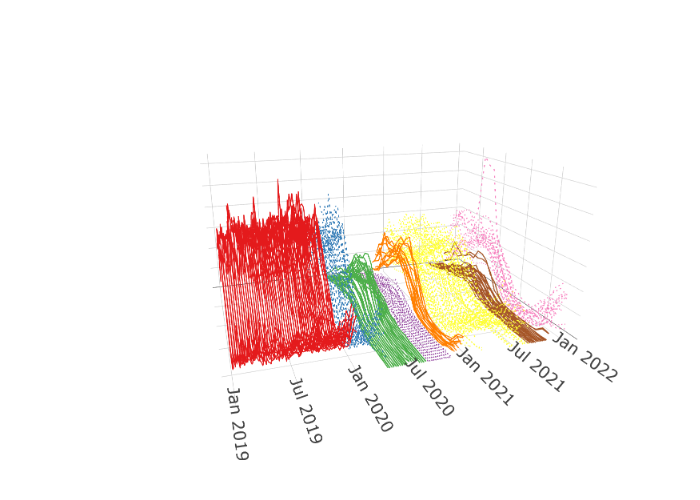}
    \caption{{\bf Segmented Melbourne pedestrian counts.} Segmented hourly pedestrian counts on a street in Melbourne Australia (2019 - 2021). Colors and line types indicate approximately homogeneous regions defined by change points detected using MST-$15$ and the max-type test statistic in \eqref{test-def:4} with $L^2$ distance.}
    \label{fig:Australia_changes}
\end{figure}

Figure \ref{fig:Australia_changes} shows approximately homogeneous regions demarcated by the $7$ detected changes with colored segments with differing line types. The detected changes seem to match the apparent distributional changes even before considering actual events. The first change is detected on 22 December 2019 ($p<0.01$), which shows a change of max daily pedestrian counts, perhaps due to the seasonal change. The next change is detected on 21 March 2020 ($p<0.01$), near the onset of COVID-19 and related lockdowns in Melbourne. Several changes follow throughout the year, mirroring the easing and strengthening of restrictions throughout the city. The change on 2 August 2020 ($p<0.01$), is near another lockdown with the onset of additional variants, followed by a change on 25 October 2020 ($p<0.01$). This October change is near the end of another round of lockdowns, at which point Melbourne had been under lockdown longer than any other city. The change detected on 6 December 2020 ($p=0.02$) was at time where people were excited to enjoy the summer weather. A change is detected on 7 August 2021 ($p=0.01$), near another few short lockdowns in the Australian winter. The final change is detected on 24 October 2021 ($p<0.01$) as people again walk the streets in the summer.

\subsection{Twitter stock returns}

\begin{figure}[th]
    \centering
    \includegraphics[trim=7cm 1cm 2.5cm 5.5cm,width=0.6\textwidth,clip]{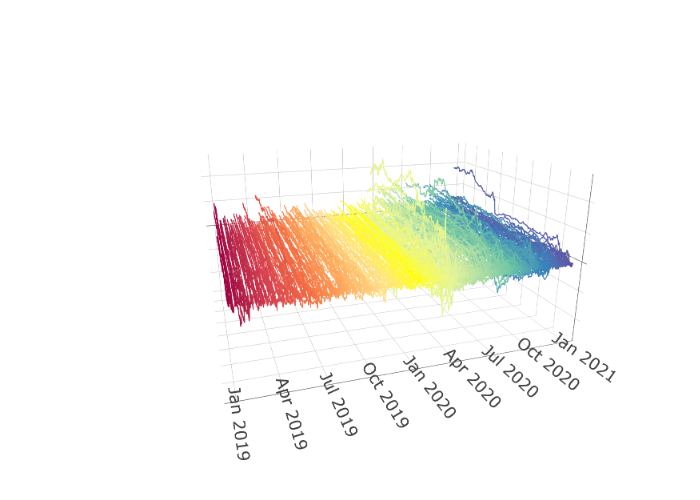}
    \caption{{\bf Twitter CIDRs.} Twitter daily CIDRs (2019 - 2020), recorded at a one-minute resolution.}
    \label{fig:twitter}
\end{figure}

Stock prices are commonly thought of as FTS, due to the high frequency of trades and constant availability of prices. In this section, the now de-listed Twitter\footnote{X, after acquisition by Elon Musk on 28 October 2022} stock data are considered. The data were recorded for every minute of trading hours from January 2019 to the end of December 2020. However, instead of directly studying the price data (which likely contains trends), intraday cumulative returns (CIDRs) are used \citep{gabrys:horvath:kokoszka:2010}. 

If $P_i(t_j)$, $i=1,\dots,n$, $j=1,\dots,m$ is the price of a financial asset at time $t_j$ on day $n$, then the CIDR curves are defined as
\begin{equation*}
    r_i(t_j) = 100 \left[\log P_i(t_j)-\log P_i(t_1) \right], \quad j=1,\dots, m, \quad i=1,\dots,n\,.
\end{equation*}
The CIDRs standardize the returns and ensure the starting value of each day is zero. Changes in mean are widely removed, and higher order changes are of primary interest in the CIDRs. The Twitter CIDRs are shown in Figure \ref{fig:twitter}.

\begin{figure}[th]
    \centering
    \includegraphics[trim=7cm 1cm 2.5cm 5.5cm,width=0.6\textwidth,clip]{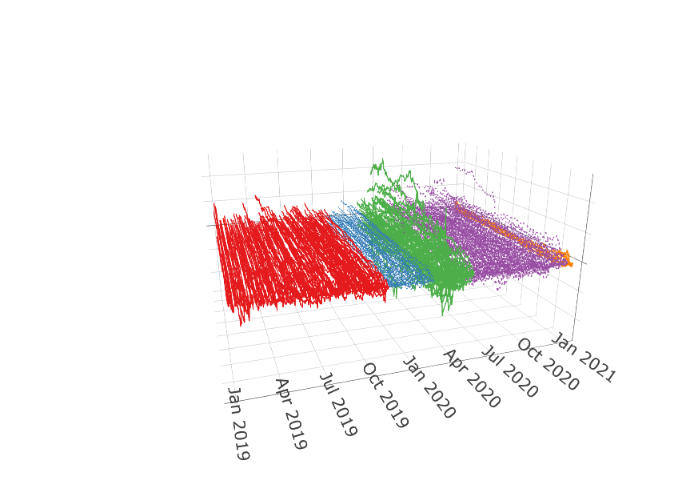}
    \caption{{\bf Segmented Twitter CIDRs.} Segmented daily Twitter CIDRs (2019 - 2020), recorded at one-minute resolution. Approximately homogeneous regions are indicated by colors and line types. The regions are demarcated by estimated changes from the max-type graph-based test statistic defined in \eqref{test-def:4} using a MST-$15$ graph and $L^2$ distance.}
    \label{fig:twitter_changes}
\end{figure}

Approximately homogeneous regions in the daily CIDR Twitter data are colored and given different line types in Figure \ref{fig:twitter_changes}. The detected changes align well with known stock market events. The first change is on 31 October 2019 ($p=0.01$), which starts a period of relative stability and corresponds to a stock market surge and the lowering of the US federal funds rate. The next change on 4 February 2020 ($p<0.01$) aligns with stock market crash and early instability due to the onset of coronavirus 2019 (COVID-19) in the US. Likewise the change on 4 May 2020 ($p<0.01$) relates to the early lockdowns and continued instability of COVID-19. Both periods are defined by volatility reflecting investor sentiment and market conditions related to lock-down and work-from-home policies. The final change is on 17 December 2020 ($p<0.01$), around the time that the federal funds rate reached its lowest.

\subsection{Spanish electricity prices} \label{sec:graph_electricity}

\begin{figure}[th]
    \centering
    \begin{subfigure}{0.45\textwidth}
        \includegraphics[trim=7cm 1cm 2.5cm 5.5cm,width=\textwidth,clip]{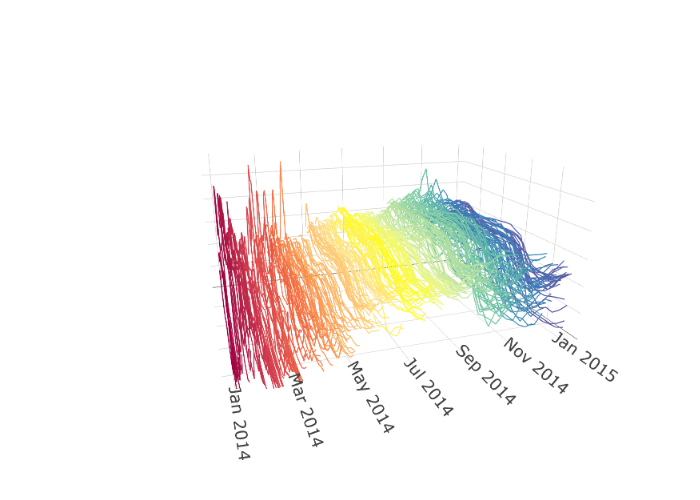}
        \caption{Electricity prices}
        \label{fig:graph_electricity}
    \end{subfigure}
    \begin{subfigure}{0.45\textwidth}
        \includegraphics[trim=7cm 1cm 2.5cm 5.5cm,width=\textwidth,clip]{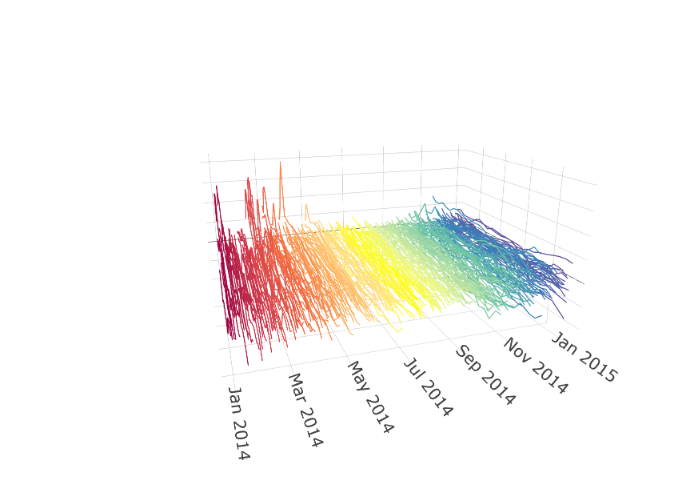}
        \caption{Residuals of the $HU(7)$ model}
        \label{fig:graph_electricity_residuals}
    \end{subfigure}
    \caption{{\bf Electricity and its model residuals.} Hourly electricity prices in 2014 for Spain and the $HU(7)$ electricity model residuals. (a) The hourly electricity prices. (b) The residuals of the electricity prices when modeled using the $HU(7)$ model with a Holt-Winters exponential smoothing model and weekly seasonality.}
\end{figure}

Many FTS exhibit strong dependence between observations. \cite{gonzalez:munoz:alonso:2018}, \cite{mestre:portela:rice:munoz:2021}, and \cite{horvath:rice:vanderdoes:2025} consider hourly electricity prices for Spain in 2014. Such data exhibits natural trends; see Figure \ref{fig:graph_electricity}. \cite{hyndman:ullah:2007} propose forecasting FTS through modeling components of functional principal components analysis. We term this approach $HU(J)$. The approach states that if data can be well-approximated by its projection onto $J$ orthonormal basis functions then
\begin{align*}
    X_i(t) \approx \hat{\mu}(t) + \sum_{j=1}^J {\xi}_{i, j} \hat{\phi}_{j}(t), \;\; t\in [0,1]\,,
\end{align*}
where the $\hat{\mu}(\cdot)$ is the sample mean function, ${\xi}_{i,j}$ is the principal component score, and $\hat{\phi}_{j}(\cdot)$ is the eigenfunction of the sample covariance operator.

\begin{figure}[th]
    \centering
    \includegraphics[trim=7cm 1cm 2.5cm 5.5cm,width=0.6\textwidth,clip]{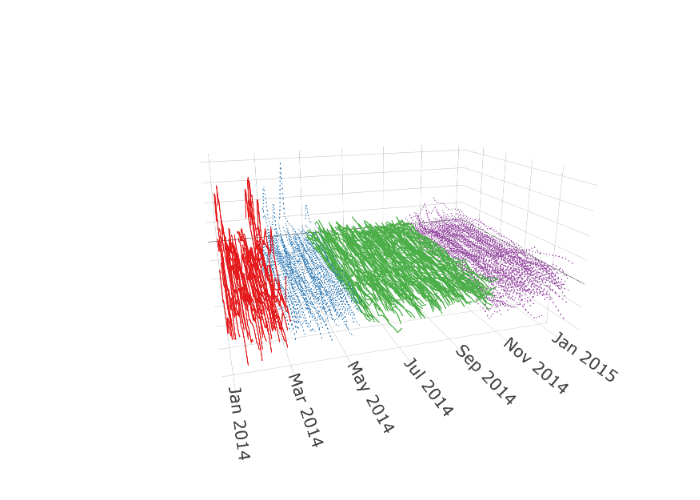}
    \caption{{\bf Segmented electricity residuals.} Segmentation for the residuals of the $HU(7)$ electricity model based on max-type graph-based change point detection using MST-$15$ and $L^2$ distance. Regions of approximate homogeneity are indicated by line type and line color.}
    \label{fig:elec_resid_changes}
\end{figure}

The $HU(J)$ model first requires approximation of the data, where $J$ is selected according to some criteria. We used total variance explained (TVE); see \cite{ramsay:silverman:2005}. We explored several values for TVE and observed little difference. Here we use $J=7$, which explained 99\% of the TVE, similar to that of \cite{horvath:rice:vanderdoes:2025}. On each score series ${\xi}_{i,j}, i=1,...,n$ for $j=1,\dots,J$, a scalar time series model is fit. We used Holt-Winters exponential smoothing with a weekly seasonality fit as implemented in the {\tt forecast} package in {\tt R}; see \cite{hyndman:khandakar:2008} for details. The $HU(J)$ model constructs the functional forecasts and fitted values as $\hat{X}_i(t) = \hat \mu(t) + \sum_{j=1}^J \hat{{\xi}}_{i,j} \hat{\phi}_j(t)$. Residual curves for the data can be computed as $r_i(t) = X_i(t) - \hat{X_i}(t)$. Following this, the residuals curves for the electricity prices are shown in Figure~\ref{fig:graph_electricity_residuals}.

Three change were detected on the $HU(7)$ model residuals. These changes generally align with the seasons and are detected on the dates February 24 ($p=0.01$), May 4 ($p<0.01$), and September 30 ($p<0.01$). The segmentation of the residuals is indicated by different colors and line types in Figure~\ref{fig:elec_resid_changes}. The residuals exhibit significantly more volatility in the winter models than in the summer months. These changes are similar to those detected in previous works, e.g. \cite{horvath:rice:vanderdoes:2025}.

\section{Discussion} \label{sec:graph_conclusion}

Graph-based change point detection is an active area of research, and a valuable addition to the toolbox of functional time series analysis. Graph-based methods exhibited strong versatility and may be of valuable in many fields. In simulations, graph-based change point detection often provided significant improvement over existing functional methods. Even in low sample sizes, graph-based methods demonstrated sensitivity to various types of changes. They are also non-parametric and robust to violations of assumptions. Graph-based change point detection performed well for a number of tuning parameters, i.e. tree type, number of orthogonal trees, test statistics, and distance metric.

In this paper, we offered several reasonable ``default'' tuning parameter choices and considerations when selecting these parameters. Even when the tuning parameters are not chosen optimally, the graph-based methods often exhibited power in a variety of scenarios. The graph-based methods returned reasonable segmentations when applied to multi-year pedestrian counts, high-resolution stock returns, and continuous electricity prices.

We hope graph-based methods will be more commonly used for functional time series analysis and encourage further investigation. Graph-based methods are often computational fast and may find value in fields were data sizes are large. We also encourage examination into approximations for graphs and $p$-values to further improve on the computational efficiency of the methods. Investigations into dependent data and confidence intervals for change points detected via graph-based change point detection would also be valuable contributions.





\bibliographystyle{elsarticle-harv} 
\bibliography{refs}

\end{document}